\documentstyle[12pt,epsfig,a4]{article} 
\newcommand{\vs}{\vspace{-0.25cm}}

\newcommand{\vlk}{$V_{{\rm low}-k}$ }

\newcommand{\be}{\begin{equation}}
\newcommand{\ee}{\end{equation}}

\begin{document}

\begin{center}
\large{\bf Density-dependent nuclear interactions and the beta decay of $^{14}$C:\\ 
chiral three-nucleon forces and Brown-Rho scaling}\footnote{Work supported in 
part by BMBF, GSI and by the DFG cluster of excellence: Origin and Structure 
of the Universe.}

\bigskip 

J.\ W.\ Holt, N.\ Kaiser, and W.\ Weise\\

\bigskip

\small{Physik Department, Technische Universit\"{a}t M\"{u}nchen,\\
    D-85747 Garching, Germany}

\end{center}

\bigskip

\begin{abstract}
We study the role of density-dependent low-momentum nucleon-nucleon 
interactions in describing the anomalously long beta decay lifetime of 
$^{14}$C. We approach this problem both from the perspective of chiral 
effective field theory, in which genuine three-body forces generate an
effective density-dependent two-body interaction, as well as from the 
perspective of Brown-Rho scaling, in which the masses and form factor cutoffs
in one-boson-exchange interactions are modified in a dense nuclear medium 
due to the partial restoration of chiral symmetry. The beta decay transition 
of $^{14}$C to the ground state of $^{14}$N is calculated within 
the shell model using a model space consisting of two $0p$-shell holes within
a closed $^{16}$O core. The effective $0p$-shell interaction is calculated
up to second order in perturbation theory with single-particle energies
extracted from experiment. We find that both
three-nucleon forces and Brown-Rho scaling medium modifications give
qualitatively similar results not only for the ground state to ground state
Gamow-Teller transition but also for Gamow-Teller transitions from excited
states of $^{14}$C to the ground state of $^{14}$N. In this way, it is
observed that at a low-momentum scale of $\Lambda_{\rm low-k}=2.1$ fm$^{-1}$, 
medium-modifications of the nuclear force play an essential 
role in increasing the lifetime of $^{14}$C from a few minutes to an 
archaeologically long one of 5730 years.
\end{abstract}

\bigskip
{\small PACS: 21.30.Fe, 21.60.Cs, 23.40.-s}

\section{Introduction}
In recent years our understanding of nuclear structure physics, and
the nuclear force in particular, has benefited greatly by incorporating 
aspects of the fundamental theory of strong interactions, QCD. Although
QCD is highly nonperturbative at the energy scales 
characteristic for nucleons bound within nuclei, one can exploit the symmetry
breaking pattern of QCD, namely the explicit and spontaneous breaking of chiral
symmetry, to constrain the underlying effective theory for interacting 
nucleons. At low energies, this is accomplished through chiral 
effective field theory \cite{weinberg90}--\cite{epelbaum06},
in which the light pseudoscalar Nambu-Goldstone bosons (e.g.\ pions) 
generated by the spontaneous breaking of chiral symmetry comprise the 
low-energy mesonic degrees of freedom relevant for nuclear interactions. All other
heavier mesons are integrated out and their effects are absorbed in a series of 
short-range contact interactions.

The above description utilizes the spontaneous breaking of chiral symmetry 
at low energies. However, as the temperature and/or density of a hadronic
medium increases, the order parameter for chiral symmetry breaking (the scalar
quark condensate $\langle \bar{q} q \rangle$) decreases in magnitude and is 
expected to vanish at an energy density around 1 GeV/fm$^3$, signaling the 
restoration of chiral symmetry. The vacuum structure of QCD, encoded in various
quark and gluon condensates, is expected to play an important role in determining
the properties (e.g.\ masses) of light hadrons \cite{adami91}--\cite{hatsuda92}, 
with the scalar quark condensate being of primary importance. The work
of Brown and Rho \cite{brownrho91,brownrho96} was one of the first attempts 
to connect medium-dependent 
meson masses to the partial restoration of chiral symmetry as the temperature 
and/or density changes. By imposing scale invariance -- a symmetry of classical, 
massless QCD -- on chiral effective Lagrangians, Brown and Rho 
obtained a scaling law in which the masses of hadrons, except the pion,
decrease with the cube root of the scalar quark condensate.

A natural testing ground for Brown-Rho scaling is
heavy ion collisions, which produce hot/dense hadronic matter in the vicinity
of chiral restoration. For a recent review see refs.\ \cite{rapp00,rapp09}.
However, even for densities at the center of heavy nuclei, the scalar
quark condensate is expected to decrease in magnitude by about 30\% as 
found from model
studies \cite{lutz92}, low-density expansions \cite{druk90}, and calculations
within in-medium chiral effective field theory \cite{dehomont}.
The decreasing of 
meson masses at these densities would have important consequences for nuclear 
structure, since the intermediate-range attraction and strong short-distance
repulsion in one-boson-exchange models of the nucleon-nucleon potential arise
largely from the exchange of 
the $\sigma$ and $\omega$ mesons, respectively. Moreover, phenomena that
are sensitive to the tensor component of the nucleon-nucleon interaction
would be affected by the decreasing of the $\rho$ meson mass. It was therefore
hoped that evidence for the partial restoration of chiral symmetry
at finite density could already be found by studying the properties of normal nuclei.

Brown and Rho proposed employing the hidden local symmetry 
approach \cite{bando88,harada03} as a framework for studying mass scaling effects.
In this formalism vector mesons, such as the $\rho$ and $\omega$, 
are treated as gauge bosons that acquire mass through the Higgs mechanism.
Presently a consistent hidden local symmetry approach incorporating baryons 
has yet to be fully developed. This limits the applicability of the 
formalism to dense nuclear systems. A more practical approach has been 
to directly modify the masses of mesons in one-boson exchange 
interactions \cite{brown90}--\cite{siu09}. 
This leads to a density-dependent nucleon-nucleon
interaction that can be used in nuclear structure studies. One aims to 
achieve a qualitative understanding of the
effects of in-medium hadron masses on nuclear structure.

Not only can the mass of a hadron be affected by a background medium, but more 
generally the entire spectral function will be modified due to Pauli blocking 
effects, collisional broadening, and the opening of additional decay channels in
a hadronic medium \cite{leupold10}--\cite{rappwam00}. 
Indeed the study of spectral functions has become the dominant 
focus both experimentally and theoretically for probing the tendency towards chiral
symmetry restoration below the transition point. The importance of resonance-hole 
excitations in modifying the spectral features of the vector mesons $\rho$ and 
$\omega$ has been studied in detail. 

Chiral effective field theory, representing the interface between QCD and nuclear
physics at low energy and momentum scales, takes a different approach by integrating
out mesons much heavier than pions. The short-distance dynamics is encoded in contact
terms. Intermediate-range effects that would emerge from $\sigma$ exchange and the
$\rho$-exchange tensor force, are now generated by explicit two-pion exchange 
processes, subject to Pauli principle corrections in the nuclear medium. Resonance-hole
effects are incorporated through genuine three-nucleon forces.

With the development of highly-accurate methods for solving the nuclear many-body problem 
for light nuclei, as well as infinitely-extended nuclear matter, it is now
well established that free-space two-nucleon interactions alone are insufficient
to describe the properties of dense nuclear systems. In particular, calculations using only
nonrelativistic two-body forces have been unable to accurately reproduce the
binding energies and spectra of light nuclei \cite{nogga00}--\cite{navratil07},  
the saturation binding energy and density of isospin-symmetric nuclear matter
\cite{day67}--\cite{fabroccini88}, 
and finally the nucleon-deuteron 
scattering differential cross sections at intermediate energies
\cite{sakamoto}--\cite{nemoto}. 
In all of 
these cases, the errors arising from the solution of the nuclear many-body problem 
are small compared to the uncertainties in the underlying nuclear force model,
which suggests the need to include explicitly degrees of freedom beyond nucleons and
mesons, or equivalently the need to include three-body forces.

In this work we study the unique problem of the beta decay lifetime
of $^{14}$C, which has been shown \cite{holt08,holt09} to be highly sensitive to 
density-dependent effects in the nuclear interaction. Describing the beta decay
lifetime of $^{14}$C has long been a challenge in nuclear structure physics, because
the anomalously small Gamow-Teller (GT) transition matrix element $M_{GT} \simeq 0.002$ 
results from the cancellation between terms that are otherwise on the order of 1.
The valence particles of $^{14}$C inhabit a region with a large nuclear density,
and therefore medium modifications to the nucleon-nucleon interaction could be sizable. 
In ref.\ \cite{holt08} it was suggested that a decreasing in-medium tensor force from
Brown-Rho scaling plays a significant role in the suppression of the Gamow-Teller transition.
This was inspired by early work \cite{inglis,talmi} that highlighted the sensitivity of 
the beta decay lifetime 
to the tensor component of the nuclear interaction. In contrast,
a later study including three-nucleon forces \cite{holt09} found qualitatively similar effects
that resulted from additional short-distance repulsion coming from the contact three-body
force. Within Brown-Rho-scaled nucleon-nucleon interaction models such an effect is
generated by the decreasing of the $\omega$ meson mass. We wish to revisit the problem 
of the $^{14}$C beta decay lifetime and suggest that, indeed, in both models of the
low-momentum density-dependent nucleon-nucleon interaction, additional short-distance repulsion 
generates significant suppression of the Gamow-Teller transition. 

This paper is organized as follows. In Section \ref{ceftnnint} we describe the framework
of chiral effective field theory, and how a density-dependent two-nucleon
force can be constructed from the chiral three-body interaction by
summing over the filled Fermi sea of nucleons.
In Section \ref{brsnnint} we motivate Brown-Rho scaling and study the effect 
that such hadronic medium modifications have on the nuclear force. In Section 
\ref{results} we apply both models to the problem of the extremely long 
beta decay lifetime of $^{14}$C, which was suggested 
to be particularly sensitive to the density dependence of the nuclear force. 
We highlight the important role played by the short-distance repulsion (increasing
with the nuclear density) in both approaches to the in-medium nuclear interaction.
We end with a summary and conclusions.


\section{Density-dependent nucleon-nucleon interaction from chiral effective field theory}
\label{ceftnnint}
Chiral effective field theory has proven to be a highly useful
framework for studying the nuclear interaction problem (for current reviews, 
see refs.\ \cite{epelbaum06,epelbaum10}). By implementing the spontaneous and
explicit breaking
of chiral symmetry in QCD, one constructs a low-energy effective field theory of
interacting pions and nucleons order-by-order in powers of $Q/\Lambda_\chi$,
where $Q$ represents a nucleon momentum or the pion mass. The long- and 
intermediate-range parts
of the nuclear force are generated by $1\pi$, $2\pi$ (and more) exchange processes, 
whereas the short-distance dynamics due to heavy mesons and baryon resonances
are integrated out and their effects are encoded in nucleon-nucleon contact 
terms. Currently the nuclear interaction has been calculated up to 4th order 
(N$^3$LO) in the chiral power counting \cite{kaiser01a}--\cite{entem03}.  
It is at this order (and beyond) that it is possible to achieve an agreement 
with empirical NN scattering phase shifts that
is comparable to previous high-precision NN potentials 
\cite{cdbonn}--\cite{argonne}. 
By adjusting the 29 parameters (mostly 
low-energy constants) that occur at this order, the 1999 database for $np$ and
$pp$ elastic scattering up to $E_{\rm lab} = 290$ MeV can be fit with a 
$\chi^2$/dof of 1.1 for $np$ scattering and 1.5 for $pp$ scattering. 
Furthermore, the experimental deuteron binding energy, charge radius, and electric 
quadrupole moment are very well reproduced by the chiral N$^3$LO potential \cite{entem03}.

\begin{figure}
\begin{center}
\includegraphics[height=4.75cm]{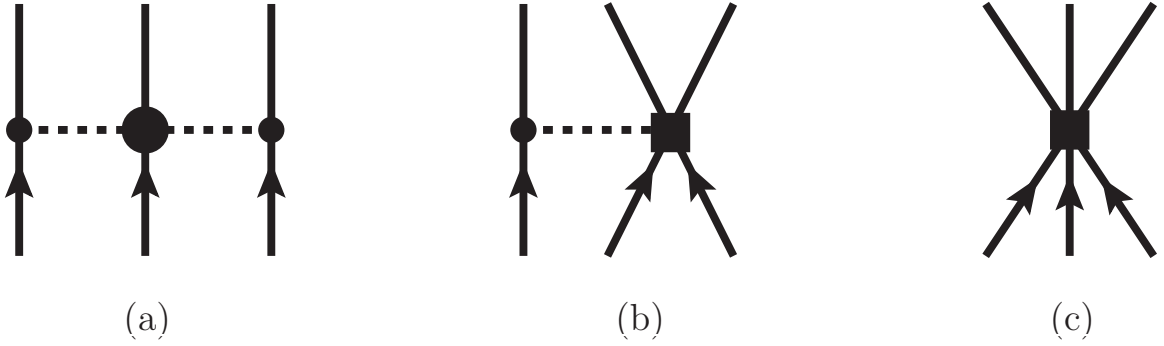}
\end{center}
\vspace{-.5cm}
\caption{The leading-order contributions to the chiral three-nucleon force: (a)
the long-range $2\pi$-exchange force $V_{3N}^{(2\pi)}$, (b) the medium-range 
$1\pi$-exchange force $V_{3N}^{(1\pi)}$, and (c) the short-range contact interaction
$V_{3N}^{(\rm ct)}$.}
\label{tnff}
\end{figure}

One decisive practical advantage of the chiral effective field theory approach is that
two- and many-body forces are generated consistently within the same 
framework. Chiral three-nucleon forces first appear at N$^2$LO. 
There are three different components, which we show in diagrams (a), (b), and
(c) in Fig.\ \ref{tnff}. The long-range, two-pion exchange interaction is given by
\begin{equation}
V_{3N}^{(2\pi)} = \sum_{i\neq j\neq k} \frac{g_A^2}{8f_\pi^4} 
\frac{\vec{\sigma}_i \cdot \vec{q}_i \, \vec{\sigma}_j \cdot
\vec{q}_j}{(\vec{q_i}^2 + m_\pi^2)(\vec{q_j}^2+m_\pi^2)}
F_{ijk}^{\alpha \beta}\tau_i^\alpha \tau_j^\beta,
\label{3n1}
\end{equation}
where $g_A=1.27$, $f_\pi = 92.4$ MeV, $m_{\pi} = 138.04$ MeV is the 
average pion mass, $\vec{q}_i=\vec{p_i}^\prime -\vec{p}_i$ is the difference 
between the final and initial momentum of nucleon $i$, and
\begin{equation}
F_{ijk}^{\alpha \beta} = \delta^{\alpha \beta}\left (-4c_1m_\pi^2
 + 2c_3 \vec{q}_i \cdot \vec{q}_j \right ) + 
c_4 \epsilon^{\alpha \beta \gamma} \tau_k^\gamma \vec{\sigma}_k
\cdot \left ( \vec{q}_i \times \vec{q}_j \right ).
\label{3n4}
\end{equation}
The three low-energy constants of the long-range 3NF, namely 
$c_1 =-0.76\,$GeV$^{-1}$, $c_3=-4.78\,$GeV$^{-1}$, and $c_4 =3.96\,$GeV$^{-1}$
appear in the two-pion exchange component of the NN interaction and therefore 
can be determined from fits to low-energy NN phase shifts \cite{rentmeester}. 
The medium-range, one-pion exchange three-nucleon interaction is proportional
to the low-energy constant $c_D/\Lambda_\chi$ and has the analytic form
\begin{equation}
V_{3N}^{(1\pi)} = -\sum_{i\neq j\neq k}
\frac{g_A c_D}{8f_\pi^4 \Lambda_\chi} \frac{\vec{\sigma}_j \cdot \vec{q}_j}{\vec{q_j}^2+m_\pi^2}
\vec{\sigma}_i \cdot
\vec{q}_j \, {\vec \tau}_i \cdot {\vec \tau}_j ,
\label{3n2}
\end{equation}
with $\Lambda_{\chi} = 700$ MeV. The filled black square in diagram (b)
of Fig.\ \ref{tnff} symbolizes the corresponding 
two-nucleon one-pion contact interaction. Finally, the short-distance contact three-body
force is proportional to the low-energy constant $c_E/\Lambda_\chi$ and is given by
\begin{equation}
V_{3N}^{(\rm ct)} = \sum_{i\neq j\neq k} \frac{c_E}{2f_\pi^4 \Lambda_\chi}
{\vec \tau}_i \cdot {\vec \tau}_j.
\label{3n3}
\end{equation}
In eqs.\ (\ref{3n1})-(\ref{3n3}) the nucleon labels $i,j,k$ can take 
the values 1,2,3 which results in six possible permutations for each sum. 
The low-energy constants of the medium-range ($V_{3N}^{(1\pi)}$) and 
short-range ($V_{3N}^{(\rm ct)}$) three-body forces are fit to properties of few-nucleon
systems, such as the triton binding energy together with the $nd$ doublet 
scattering length \cite{epelbaum02}, the $^4$He binding energy \cite{nogga06},
or the binding energies and spectra of light nuclei \cite{navratil07}.
The N$^3$LO three-body interaction contains numerous one-loop diagrams but
no additional low-energy constants and is currently under construction \cite{epelbaum10}.

In this work we utilize renormalization group techniques to decimate the 
chiral N$^3$LO two-body interaction down to an energy scale in the vicinity
of $\Lambda = 2.0$ fm$^{-1}$, following the details provided in refs.\ 
\cite{bogner02,bogner03}.
Such a decimation procedure has the desirable feature that the resulting 
``low-momentum interaction'', referred to as $V_{\rm low-k}$, is nearly 
unique regardless of which underlying high-precision NN potential is employed, 
and therefore $V_{\rm low-k}$ provides a model-independent description of 
the nucleon-nucleon interaction below $\Lambda=2.0$ fm$^{-1}$. The analogous 
evolution of 
realistic three-body forces down to this scale has yet to be achieved. 
However, since such a decimation would change primarily
the short-distance part of the three-body interaction, a practical alternative
has been to treat the low-energy constants $c_D$ and $c_E$ as functions of the
variable $\Lambda_{\rm low-k}$. By fitting $c_D(\Lambda_{\rm low-k})$ and 
$c_E(\Lambda_{\rm low-k})$ to the binding energies of $^3$H and $^4$He, one 
obtains \cite{nogga} the values shown in Table \ref{cdcelowk} for two different
choices of the low-momentum cutoff.

\begin{table}[htb]
\begin{center}
\begin{tabular}{|c||c|c|} \hline
$\Lambda_{\rm low-k}$ & $c_D$ & $c_E$\\ \hline
2.1 fm$^{-1}$ & $-2.062$ & $-0.625$\\ \hline
2.3 fm$^{-1}$ & $-2.785$ & $-0.822$\\ \hline
\end{tabular}
\caption{The values of the low-energy constants $c_D$ and $c_E$ of the chiral 
three-nucleon interaction fit to the binding energies of $A=3,4$ nuclei for 
different values of the momentum cutoff $\Lambda_{\rm low-k}$.}
\label{cdcelowk}
\end{center}
\end{table}

We now show how to derive from the leading-order chiral three-nucleon 
interaction, eqs.\ (\ref{3n1}--\ref{3n3}), an effective density-dependent 
in-medium NN interaction. We keep our discussion brief and refer the interested
reader to the original references \cite{holt09,holt10}. 
We consider the on-shell scattering of two nucleons in isospin-symmetric 
(spin-saturated) nuclear
matter of density $n=2k_f^3/3\pi^2$ in the center-of-mass 
frame, $N_1(\vec p\,)+ N_2(-\vec p\,) \to N_1(\vec 
p+\vec q\,) + N_2(-\vec p-\vec q\,)$, where $k_f$ is the Fermi momentum. 
The magnitude of the in- and out-going nucleon momenta is $|\vec p\,| = p 
= |\vec p+\vec q\,|$, and $q= |\vec q\,|$ is the magnitude of 
the momentum transfer. The resulting interaction in momentum-space has the
following (general) form
\begin{eqnarray}
V(\vec p,\vec q) &=& V_C + \vec \tau_1 \cdot \vec \tau_2\, W_C + 
\left [V_S + \vec \tau_1 \cdot \vec \tau_2 \, W_S \right ] \vec \sigma_1 \cdot 
\vec \sigma_2 + \left [ V_T + \vec \tau_1 \cdot \vec \tau_2 \, W_T \right ] 
\vec \sigma_1 \cdot \vec q \, \vec \sigma_2 \cdot \vec q \nonumber \\
&+& \left [ V_{SO} + \vec \tau_1 \cdot \vec \tau_2 \, W_{SO} \right ] \,
i (\vec \sigma_1 + \vec \sigma_2 ) \cdot (\vec q \times \vec p) \nonumber \\
&+& \left [ V_{Q} + \vec \tau_1 \cdot \vec \tau_2 \, W_{Q} \right ] \,
\vec \sigma_1 \cdot (\vec q \times \vec p)\, \vec \sigma_2 \cdot (\vec q \times
\vec p).
\end{eqnarray}
The subscripts refer to the central (C), spin-spin (S), tensor (T), 
spin-orbit (SO), and quadratic spin orbit (Q) components of the NN 
interaction, each of which occurs in an isoscalar (V) and an isovector (W) version.

\begin{figure}
\begin{center}
\includegraphics[scale=1,clip]{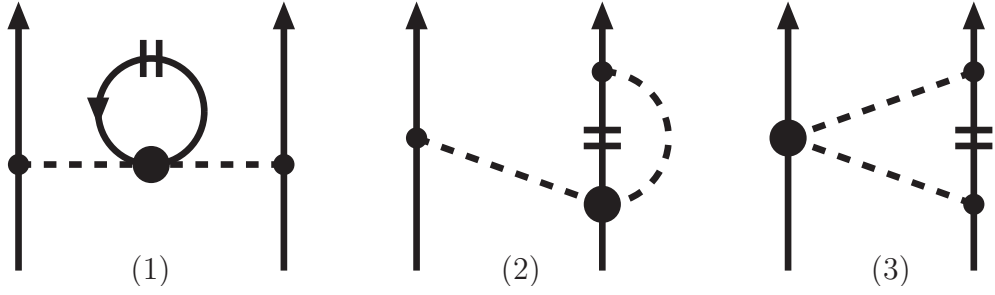}
\end{center}
\vspace{-.5cm}
\caption{In-medium NN interaction generated by the two-pion exchange component
of the chiral three-nucleon interaction. The short 
double-line symbolizes the filled Fermi sea of nucleons. 
Reflected diagrams of (2) and (3) are not shown.}  
\label{mfig1}
\end{figure}

\begin{figure}
\begin{center}
\includegraphics[scale=1,clip]{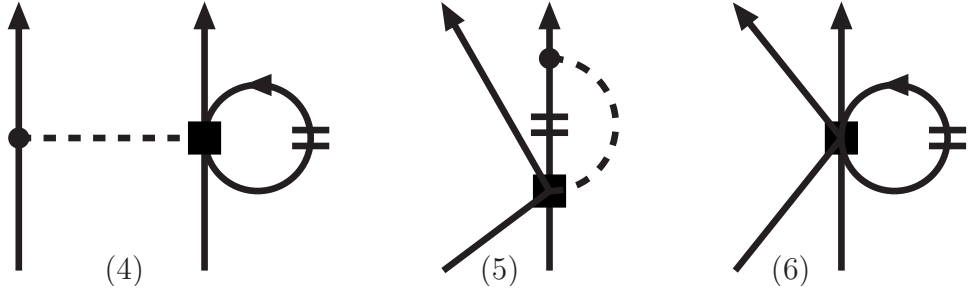}
\end{center}
\vspace{-.5cm}
\caption{In-medium NN interaction generated by the one-pion exchange component
and the short-range component of the chiral three-nucleon interaction.} 
\label{mfig2}
\end{figure}

In Figs.\ \ref{mfig1} and \ref{mfig2} we show the six topologically distinct 
contributions to the in-medium density-dependent nucleon-nucleon interaction
to one-loop order. The long-range two-pion exchange three-nucleon force
generates the diagrams shown in Fig.\ \ref{mfig1}. Tadpole diagrams 
obtained by contracting the in- and outgoing lines at a $\pi N N$ vertex vanish, which
leaves only three topologically distinct nonvanishing contributions. The short double-line on 
a nucleon propagator symbolizes the filled Fermi sea of nucleons, which introduces
the ``medium insertion'' $-2\pi \delta(k_0)\,\theta(k_f- |\vec k\,|)$ in the in-medium 
nucleon propagator. Diagram (1) in Fig.\ \ref{mfig1} represents one-pion exchange
with a Pauli-blocked pion self-energy correction. The intermediate nucleon momentum
is independent of the momentum transfer $\vec{q}$ carried by the pions and therefore
the integration over the filled Fermi sea of nucleons produces a correction 
proportional to the density $n$:
\begin{equation} V_{NN}^{\rm med,1}= {g_A^2 M_N n \over 8 \pi f_\pi^4}\,\vec \tau_1  
\cdot \vec \tau_2 \,{\vec \sigma_1 \cdot\vec q \,\vec \sigma_2 \cdot \vec q 
\over (m_\pi^2 + q^2)^2}\,(2c_1 m_\pi^2 +c_3 q^2)\,.
\label{med1}
\end{equation}
Since $c_{1,3}<0$, this term gives a (large) enhancement of the bare $1\pi$-exchange:
\begin{equation} V_{NN}^{(1\pi)} = - {g_A^2 M_N \over 16 \pi f_\pi^2} \vec \tau_1 \cdot  
\vec \tau_2 \,  {\vec \sigma_1 \cdot \vec q \,\vec \sigma_2 \cdot \vec q\over  
m_\pi^2  + q^2}\,. 
\label{ope}
\end{equation}   
In part this enhancement can be interpreted in terms of the reduced in-medium 
pion decay constant, 
$f_{\pi,s}^{*2}=f_\pi^2+2c_3\rho$, which is associated with 
the spatial components of the axial current. 
Diagram (2) of Fig.\ \ref{mfig1} represents a vertex correction to one-pion exchange.
Fermi sphere integrals, $\int_{|\vec{k}|\leq k_F} d^3k$, over a 
single static pion propagator give rise to $p$-dependent
functions $\Gamma_j(p)$, where the dependence on $k_F$ is implicit. For the sake of
brevity and continuity in the current discussion, we have left the explicit expressions
for the $\Gamma_j(p)$ in the Appendix. Summing diagram (2) of Fig.\ \ref{med1} together with
the three reflected diagrams yields the following contribution
\begin{eqnarray} V_{NN}^{\rm med,2}&=& {g_A^2 M_N \over 32\pi^3 f_\pi^4}\vec \tau_1  
\cdot \vec \tau_2 \,  {\vec \sigma_1 \cdot \vec q \,\vec \sigma_2 \cdot \vec q 
\over m_\pi^2 + q^2}\, \bigg\{-4c_1 m_\pi^2 \Big[\Gamma_0(p)+\Gamma_1(p) \Big]
\nonumber \\ && - (c_3+c_4) \Big[q^2 \Big(\Gamma_0(p)+2\Gamma_1(p)+
\Gamma_3(p)\Big)+ 4\Gamma_2(p)\Big] + 4c_4 \bigg[ {2k_f^3 \over
  3}-m_\pi^2\Gamma_0(p)\bigg] \bigg\}\,.
\label{med2}
\end{eqnarray}
By analyzing the momentum- and density-dependent factors in eq.\ (\ref{med2}) relative to
$V_{NN}^{(1\pi)}$, one finds that this contribution corresponds to a (large) reduction
of the $1\pi$-exchange in the nuclear medium. Approximately, this feature can
be interpreted in terms of a quenched nucleon axial-vector constant
$g_A^*$.

Diagram (3) in Fig.\ \ref{mfig1} represents Pauli-blocked $2\pi$-exchange. The
Fermi sphere integrals over two pion propagators give rise to $p$- and $q$-dependent 
functions $G_j(p,q)$ which we tabulate in the Appendix. Diagram (3) taken together 
with the corresponding reflected diagram then produces the contribution
\begin{eqnarray} V_{NN}^{\rm med,3} &=& {g_A^2 M_N \over 64 \pi^3 f_\pi^4}\bigg\{  -12 c_1 
m_\pi^2 \Big[2\Gamma_0(p)- (2m_\pi^2+q^2) G_0(p,q)\Big] \nonumber \\ && -c_3
\Big[8 k_f^3-12(2m_\pi^2+q^2) \Gamma_0(p) -6q^2\Gamma_1(p)+3(2m_\pi^2+q^2)^2 G_0(p,q)
\Big]\nonumber \\&& +4c_4\,\vec \tau_1 \cdot  \vec \tau_2\, (\vec \sigma_1
\cdot \vec \sigma_2\, q^2 - \vec \sigma_1 \cdot  \vec q \,\vec\sigma_2\cdot
\vec q\,) G_2(p,q) \nonumber \\ && -(3c_3+c_4\vec \tau_1 \cdot \vec \tau_2 )\,
i ( \vec \sigma_1 +\vec \sigma_2 )\cdot(\vec q \times \vec p\,)\Big[2\Gamma_0(p)
+2\Gamma_1(p) - (2m_\pi^2+q^2)\nonumber \\ &&\times \Big(G_0(p,q)+2 G_1(p,q)
\Big) \Big] -12 c_1 m_\pi^2\, i ( \vec \sigma_1 +\vec \sigma_2 ) \cdot(\vec q 
\times \vec p\,)  \Big[G_0(p,q)+2 G_1(p,q)\Big] \nonumber \\ && + 4c_4\, 
\vec\tau_1 \cdot \vec \tau_2 \,\vec \sigma_1 \cdot (\vec q \times \vec p\,)\,
\vec\sigma_2 \cdot( \vec q \times \vec p \,) \Big[G_0(p,q) +
4G_1(p,q)+4G_3(p,q) \Big]\bigg\}\,. 
\label{med3}
\end{eqnarray}
In comparison to the analogous $2\pi$-exchange interaction in vacuum 
(see section 4.2 in ref.\ \cite{nnpap}) the Pauli blocking in the 
nuclear medium  has generated additional spin-orbit terms, $i(\vec\sigma_1 
+\vec \sigma_2 )\cdot(\vec q \times \vec p\,)$, and quadratic spin-orbit terms, $\vec \sigma_1 
\cdot (\vec q \times \vec p\,)\,\vec\sigma_2 \cdot( \vec q \times \vec p \,)$, 
written in the last three lines of eq.\ (\ref{med3}).

Next we consider the $1\pi$-exchange component of the chiral three-nucleon
interaction proportional to the parameter $c_D/\Lambda_{\chi}$. Contracting two
external lines results in two (nonvanishing) topologies. By closing a nucleon line at the
contact vertex, one obtains a vertex correction to $1\pi$-exchange:
\begin{equation} V_{NN}^{\rm med,4} = -{g_A M_N c_D n \over 32 \pi f_\pi^4\Lambda_{\chi}}
\,\vec \tau_1 \cdot \vec \tau_2 \,{\vec \sigma_1 \cdot\vec q \,\vec \sigma_2 
\cdot \vec q  \over  m_\pi^2 + q^2} \,.
\label{med4}
\end{equation}
We note that this contribution is proportional to the nuclear density $n$ due to the 
$q$-independent Fermi sphere integral.
Since $c_D$ is negative, $V_{NN}^{\rm med,4}$ reduces the bare $1\pi$-exchange,
roughly by about $16\%$ at normal nuclear matter density $\rho_0=0.16\,
$fm$^{-3}$. Diagram (5) in Fig.\ \ref{mfig2} represents Pauli-blocked (pionic) 
vertex corrections to the short-range NN interaction. The corresponding 
contribution to the density dependent in-medium NN interaction reads:
\begin{eqnarray} V_{NN}^{\rm med,5}&=& {g_A M_N c_D\over 64 \pi^3 f_\pi^4\Lambda_{\chi}} \left \{\vec 
\tau_1 \cdot \vec \tau_2 \left [ 2 \vec \sigma_1 \cdot \vec \sigma_2\,\Gamma_2(p)
 +\left (\vec \sigma_1 \cdot \vec \sigma_2 \left( 2p^2-{q^2\over 2}\right) + \vec 
\sigma_1 \cdot \vec q\,\vec\sigma_2  \cdot \vec q\, 
\left(1-{2p^2\over q^2}\right) \right . \right . \right . \nonumber \\ &&  \left . \left . \left . -{2\over q^2}\,\vec\sigma_1 \cdot (\vec q \times 
\vec p\,)\,\vec\sigma_2 \cdot(\vec q\times \vec p \,)\right )
\Big(\Gamma_0(p)+2\Gamma_1(p)+\Gamma_3(p)\Big ) \right ]+4k_f^3-6m_\pi^2 \Gamma_0(p) \right \}
\label{med5}
\end{eqnarray}
Due to the special spin and isospin structure of this term, it gives nonvanishing 
contributions only to relative $S$-states as well as to the $^3S_1 - {^3D}_1$ mixing matrix element.

\begin{figure}
\begin{center}
\includegraphics[height=13cm,angle=270]{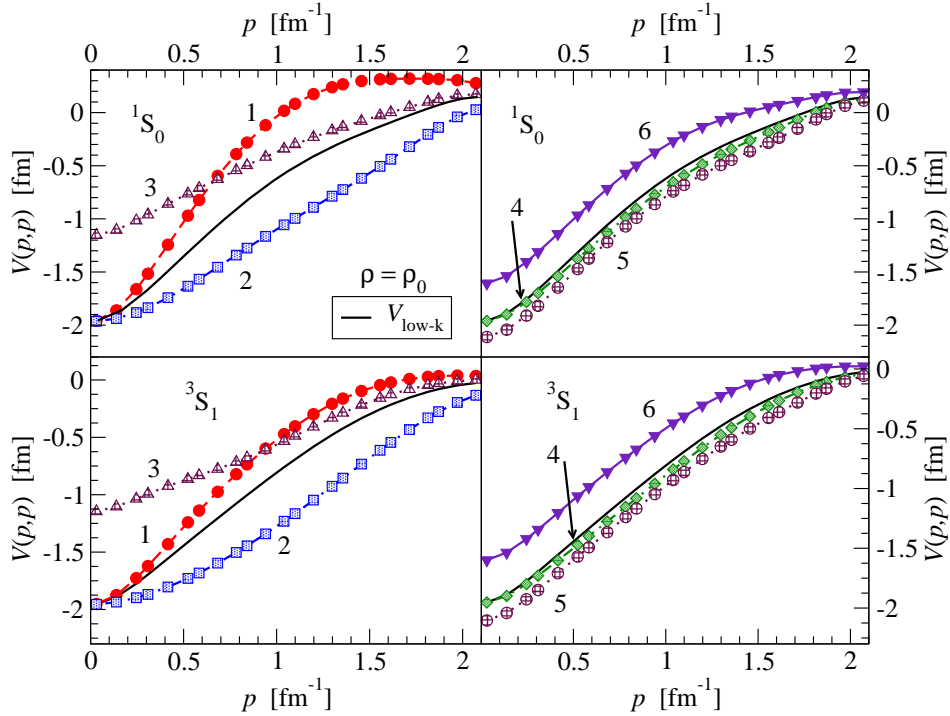}
\end{center}
\vspace{-.5cm}
\caption{Effect (at normal nuclear matter density, $n_0=0.16$ fm$^{-3}$) 
of the six different terms $V_{NN}^{\rm med,i}$ on the $S$-wave 
components of the free-space nucleon-nucleon
potential \vlk for a low-momentum cutoff of $\Lambda_{\rm low-k}=2.1$ fm$^{-1}$.}
\label{swaves}
\end{figure}

Finally, we discuss the short-range component of the chiral 3N interaction, 
represented by a three-nucleon contact-vertex proportional to $c_E/\Lambda_{\chi}$. By
closing one nucleon line (see diagram (6) in Fig.\ \ref{mfig2}) one obtains the
following contribution to the in-medium NN-interaction:
\begin{equation}  V_{NN}^{\rm med,6} =-{3 M_N c_E n \over 8 \pi f_\pi^4\Lambda_\chi} \,,
\label{med6}
\end{equation}
which simply grows linearly in density $n$ and is independent of 
spin, isospin and nucleon momenta. 

In Fig.\ \ref{swaves} we show the contribution from each of these components of the 
density-dependent NN interaction to the $^1S_0$ and $^3S_1$ diagonal partial wave matrix
elements (see ref.\ \cite{holt10} for an analysis of higher partial waves). The black solid 
line gives the matrix elements of \vlk obtained from the Idaho N$^3$LO chiral NN potential
for a momentum cutoff of $\Lambda_{\rm low-k} = 2.1$ fm$^{-1}$. The various curves
labeled $i=1,2,\dots, 6$ show the matrix elements of \vlk + $V_{NN}^{{\rm med},i}$.
We see that the effects from the long-range two-pion exchange part of the chiral
three-nucleon force are quite large. However, the pion self-energy correction 
$V_{NN}^{\rm med,1}$ largely cancels the effects of the vertex correction 
$V_{NN}^{\rm med,2}$. Since $V_{NN}^{\rm med,1}$ and $V_{NN}^{\rm med,2}$ are 
simply modifications to one-pion exchange, a similar cancellation occurs in 
all higher partial waves as well. 
The final component, $V_{NN}^{\rm med,3}$, that comes from the long-range three-body force
has a much richer spin and isospin structure, and at large distances (small momenta)
the overall effect is strongly repulsive.

The two terms $V_{NN}^{\rm med,4}$ and $V_{NN}^{\rm med,5}$ arise from the medium-range
one-pion exchange three-body force proportional to the low-energy constant $c_D$.
In Fig.\ \ref{swaves} we see that the effects are much weaker than from the 
long-range chiral three-nucleon force. Since $V_{NN}^{\rm med,4}$ is again just
a reduction of one-pion exchange, it has nonzero
contributions in all partial waves. The second term arising from the mid-range three-nucleon
force, $V_{NN}^{\rm med,5}$, is also relatively small and gives an additional (equal) attraction
in the two relative $S$-states, as well as the coupled $^3S_1- {^3D}_1$ mixing
matrix element not shown in the figure. The final contribution 
$V_{NN}^{\rm med,6}$ is a pure contact interaction and therefore contributes
only to the two relative $S$-states. The interaction is spin and isospin
independent and therefore contributes to both of these partial waves 
equally. Indeed, the additional repulsion is approximately twice as large
as the attraction coming from $V_{NN}^{\rm med,5}$, so taking these two
terms together essentially results in a weaker short-range repulsive interaction.

\begin{figure}[hbt]
\begin{center}
\includegraphics[height=15cm,angle=-90]{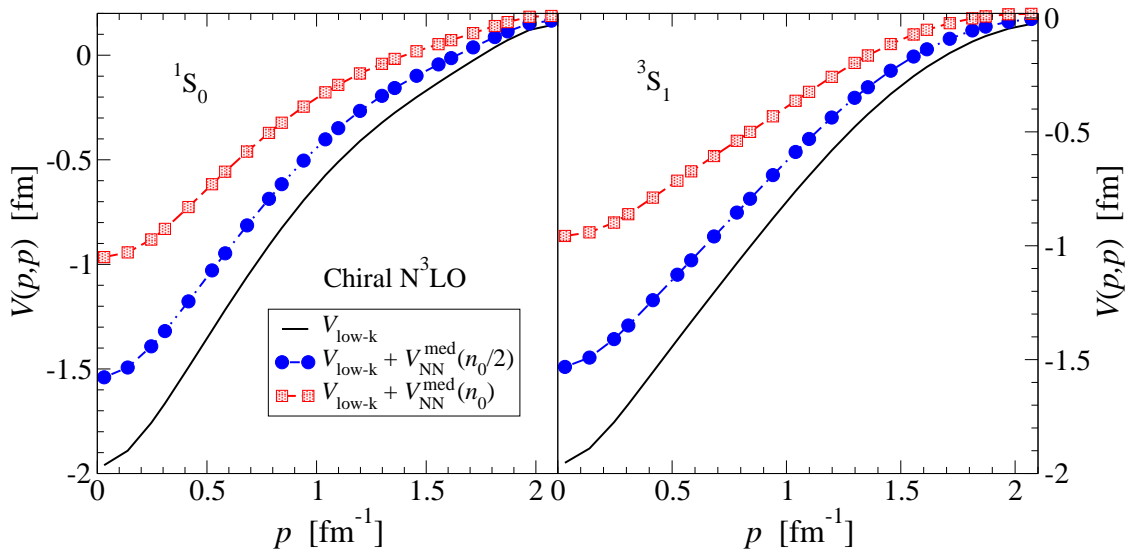}
\caption{Partial wave matrix elements for relative $S$-states from 
\vlk + $V_{NN}^{\rm med}$. The low-momentum decimation scale is taken
to be $\Lambda_{\rm low-k} = 2.1$ fm$^{-1}$.}
\label{n3lopw}
\end{center}
\end{figure}

In Fig.\ \ref{n3lopw} we show the density dependence 
(for $n=0,n_0/2,$ and $n_0$)
of the diagonal $S$-wave matrix elements of \vlk + $V_{NN}^{\rm med}$.
Overall, we find that the effects are repulsive and quite large 
in comparison to the free-space low-momentum NN interaction. 
The same holds in most higher partial waves as well, with the 
exception that the tensor component of the in-medium
interaction becomes more attractive with the density \cite{holt10}. 
The large $S$-wave repulsion in the three-body forces increases with
the density and thereby provides
the mechanism for saturation of low-momentum interactions \cite{bogner05}.
In Section \ref{results} we study the role that these density-dependent 
effects play in describing the anomalously long beta decay lifetime of $^{14}$C and
make a comparison with analogous interactions obtained from the 
Brown-Rho scaling hypothesis.

\section{Brown-Rho scaled nucleon-nucleon interactions}
\label{brsnnint}

In this section we summarize Brown-Rho scaling and describe a model
approach for implementing these effects in nuclear interactions at densities
close to that of saturated nuclear matter. The starting point is 
the scaling relation for (light) hadron masses derived by Brown and Rho \cite{brownrho91}:
\begin{equation}
\sqrt{\frac{g_A}{g_A^*}}\frac{M_N^*}{M_N} \simeq \frac{m_\sigma^*}{m_\sigma} \simeq
\frac{m_\rho^*}{m_\rho} \simeq 
\frac{m_\omega^*}{m_\omega} \simeq \frac{f_\pi^*}{f_\pi} = \Phi(n),
\label{brs0}
\end{equation}
which was obtained by implementing scale invariance in chiral effective Lagrangians
following the approach of Campbell et al.\ \cite{camp}. In eq.\ (\ref{brs0}) the
asterisk refers to an in-medium value of the given quantity, $f_\pi^*=f_{\pi,t}^*$ is
related to the time component of the axial vector current, and $\Phi$ (to
be discussed in more detail shortly) is a function of the nuclear density $n$.  
Absent from the scaling hypothesis in eq.\ (\ref{brs0}) is the pion mass, 
which in the chiral limit of massless bare quarks
would be zero due to the Goldstone boson nature of pions. The nonzero pion mass
is due rather to the explicit breaking of chiral symmetry by the quark mass term
in the QCD Lagrangian and is therefore not as sensitive to the background
medium as are the masses of other hadrons. This is supported empirically
from measurements of energy shifts in deeply bound pionic atoms \cite{suzuki04,
kolom03}.

From eq.\ (\ref{brs0}), hadronic masses
are predicted to scale with the pion decay constant $f_\pi$, the hadronic 
observable connected to the spontaneous breaking of chiral symmetry. The 
connection between $f_\pi$ and the fundamental order parameter 
$\langle \bar{q} q \rangle$ of QCD is given by the 
Gell-Mann--Oakes--Renner relation \cite{gor,lutz92}
\begin{equation}
f_\pi^2 m_\pi^2 = -m_q \langle\bar u u + \bar d d\rangle,
\end{equation}
where $m_q$ is the average (current) mass for up $u$ and down $d$ quarks. 
Assuming that the pion mass is protected by chiral invariance, this relation 
would produce
\begin{equation}
\frac{f_\pi^*}{f_\pi}= \left (\frac{\langle \bar q q\rangle^*}
{\langle\bar q q\rangle}\right )^{1/2},
\label{eqfs}
\end{equation}
where $\langle \bar{q} q \rangle^* = \langle \Omega | \bar{q} q | \Omega 
\rangle$ and $| \Omega \rangle$ denotes the ground state of the dense medium.
In the limit of low densities the scalar quark condensate is related to 
the nuclear density by \cite{druk90,coh}
\begin{equation}
\frac{\langle\bar q
q\rangle^*}{\langle\bar q q\rangle} = 1 -\frac{\sigma_{\pi
N}}{f_\pi^2 m_\pi^2}n + \cdots,
\label{eqla}
\end{equation}
where $\sigma_{\pi N}$ is the pion-nucleon sigma term that describes how the 
nucleon mass varies with the masses of bare quarks. From a dispersion relation
analysis of low-energy $\pi N$ scattering, one can deduce 
$\sigma_{\pi N} \simeq 45$ MeV \cite{gasser91}. Thus, at nuclear 
matter saturation density one would obtain from eqs.\ (\ref{brs0}), (\ref{eqfs}), and 
(\ref{eqla}) that $\Phi(n_0) \simeq 0.83$.
In Brown-Rho scaling one therefore assumes a linear decrease of hadronic
mass with the density:
\begin{equation}
\frac{m^*}{m} \simeq 1-\frac{1}{2}\frac{\sigma_{\pi
N}}{f_\pi^2 m_\pi^2}n \simeq 1-0.17\frac{n}{n_0}.
\end{equation}
Such a relationship has also be deduced from in-medium QCD sum rule analyses, which
can provide important constraints on the form the vector meson spectral
functions \cite{hatsuda92,klingl97,kwon08}.

\begin{figure}[hbt]
\begin{center}
\includegraphics[height=3.9cm]{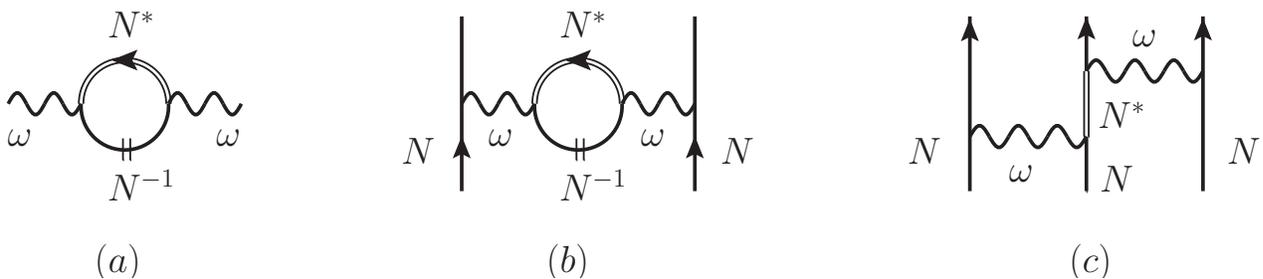}
\caption{The contribution of resonance-hole excitations to (a) the in-medium 
$\omega$ meson propagator, (b) the in-medium NN interaction resulting from
$\omega$ meson exchange, and (c) the analogous contribution to the three-nucleon force.}
\label{omegmed}
\end{center}
\end{figure}

Brown-Rho scaling was one of the first attempts to study -- through changes 
in the symmetry breaking pattern of the ground state -- the effects of a 
hot/dense medium on the properties of an embedded hadron. Subsequent work
had the broader goal to obtain the full spectral function of in-medium hadrons, 
with special emphasis on the vector mesons: $\rho$ and $\omega$. 
Within hadronic many-body calculations, self-energy corrections $\Sigma_{\rm med}$
to the meson propagator $D(\omega, \vec{q}\,)$ defined by
\begin{equation}
 D(\omega,\vec{q}\,) = \frac{1}
{\omega^2-\vec{q\,}^2-m^2-\Sigma_{\rm vac}(q)-\Sigma_{\rm med}(\omega,\vec{q}\,)}
\end{equation}
result in modifications to the spectral function $A(\omega,\vec{q}\,)$, which is proportional to the 
imaginary part of the propagator:
\begin{equation}
 A(\omega,\vec{q}\,) = \frac{1}{\pi}\frac{{\rm Im}\,\Sigma(\omega,\vec{q}\,)}
{(\omega^2-\vec{q\,}^2-m^2+{\rm Re}\,\Sigma(\omega,\vec{q}\,))^2+{\rm Im}\,
\Sigma^2(\omega,\vec{q}\,)}.
\end{equation}
The real part of the self-energy then contributes to the effective mass of the particle
while the imaginary part leads to collisional broadening of the width. 
In a background medium of cold nuclear matter, contributions to $\Sigma_{\rm med}$ 
arise from the coupling of mesons to nucleon-hole and resonance-hole excitations
\cite{peters98,klingl97,post02,friman97} as shown in Fig.\ \ref{omegmed}(a).
Calculations are simplified by invoking the following general result that holds 
at low-densities (corresponding to small Fermi momenta) and which 
relates the in-medium self-energy to the total forward scattering
amplitude $T(\vec q, \omega)$ of a vector meson on a nucleon
\begin{equation}
 \Sigma_{\rm med}(\omega,\vec{q}\,) = \int_{p_N<k_f} \frac{d^{\,3} p_N}{(2\pi)^32E_N}T(q,p_N) \simeq
\frac{n}{8m_N} \, T(\vec{q},\omega),
\end{equation}
where $n$ is the nucleon density and one assumes that the scattering
amplitude varies slowly with momentum. The result of such studies was that 
medium modifications to vector meson spectral
functions in nuclear matter are dominated by subthreshold resonances, such as the 
$D_{13}(1520)$ resonance which produces an enhancement in the low-energy 
region of the $\rho$ and $\omega$ meson spectral functions. Most calculations concluded that 
the spectral functions of vector mesons in cold nuclear matter are largely broadened
with little shift of the peak mass. However, within an in-medium QCD sum rule approach
it has been found \cite{kwon08} that the first moment of the $\rho$-meson spectral
function can 
decrease (similarly to that in eq.\ (\ref{brs0})) despite the fact that the mass peak
changes very little.

Three-nucleon forces and hadronic medium modifications provide two complementary 
(and partly equivalent) frameworks
for understanding and implementing density-dependence in the nucleon-nucleon interaction.
The role of baryonic resonances are important for each: on the one hand they give rise 
to genuine three-nucleon forces, and on the other hand they provide the main mechanism for altering
mesonic spectral functions in medium. The connection can be visualized in the three
diagrams of Fig.\ \ref{omegmed}, which show (a) a process contributing to the modification 
of the $\omega$-meson spectral function in medium, (b) the effect that such a process would
have on the one-boson-exchange NN interaction due to $\omega$ meson exchange, 
and (c) the corresponding
three-nucleon force obtained by opening the nucleon hole line in diagram (b). At present
chiral three-nucleon forces are better constrained experimentally and provide a complete 
basis for including effects of the medium, while the properties of hadrons in medium
are less constrained but have a more direct connection to questions regarding chiral 
symmetry restoration in dense nuclear systems. In the rest of this section we consider a 
specific model for implementing Brown-Rho scaling medium modifications in one-boson-exchange
interactions in order to make a quantitative comparison to models of the density-dependent
NN interaction from chiral three-nucleon forces.

The scaling relation in eq.\ (\ref{brs0}) should be regarded as only an approximate, mean-field
relation that could be altered through higher-order corrections. Indeed, it is well-established 
\cite{brown90,rapp99} that a naive 
implementation of eq.\ (\ref{brs0}) would yield far too much attraction due to the
decreasing of the effective scalar $\sigma$ meson mass. A more accurate treatment of
the intermediate-range attraction in the NN potential would be to replace the 
``ficticious'' $\sigma$ meson by a pair of correlated pions (in the $J^\pi = 0^+$ channel).
This has been done in ref.\ \cite{kim94}, where a meson exchange model for the $S$-wave 
$\pi \pi$ interaction was developed and found to be dominated by the exchange 
of the $\rho$ meson as well as repulsive $\pi \pi$ contact interactions 
constrained by chiral symmetry. With this model for the $\pi \pi$ interaction, 
the $\pi \pi \rightarrow N \bar{N}$ amplitude was then constructed, which can be related through crossing relations and a dispersion integral to the 
$ \pi N \rightarrow \pi N$ amplitude from which one can derive the effective $2\pi$-exchange
NN potential.

\setlength{\tabcolsep}{.09in}
\begin{table}[htb]
\begin{center}
\begin{tabular}{|c|c|c|c|c|} \hline
$n/n_0$ & $m_{\sigma_1}$ (MeV) & $g^2_{\sigma_1}/4\pi$ & $m_{\sigma_2}$ (MeV) & $g^2_{\sigma_2}/4\pi$ \\ \hline
0.0  & 640 & 4.36 & 377 & 0.640  \\ \hline
0.5  & 538 & 3.41 & 295 & 0.450  \\ \hline
0.75 & 471 & 2.88 & 249 & 0.309  \\ \hline
1.0  & 395 & 2.53 & 220 & 0.264  \\ \hline
1.25 & 365 & 2.35 & 200 & 0.234  \\ \hline
\end{tabular}
\caption{Masses and coupling constants for the two zero-width bosons parameterizing medium-modified
correlated $2\pi$ exchange in the Bonn-B potential. \cite{rapp99}}
\label{sigmas}
\end{center}
\end{table}

In this formalism finite density effects in the scalar-isoscalar channel 
arise through the density-depedence of the $\rho$ meson mass ($m_\rho$) 
and form factor cutoff ($\Lambda_\rho$) as well
as the $\pi \pi$ contact interactions. Moreover, the in-medium pion propagator
is dressed with $\Delta$-hole excitations. In practice, this complicated treatment of 
correlated $2\pi$ exchange was modeled with the exchange
of two scalar-isoscalar bosons with a form factor cutoff of $\Lambda_\sigma = 3$ GeV 
and masses and coupling constants shown in Table \ref{sigmas}.
Together with the scaling of the 
vector meson masses and form factor cutoffs\footnote{To the extent that the 
phenomenological form factors associated nucleon-meson interaction vertices 
are governed by the nucleon radius, they too will decrease with density due 
to the proposed increase in the nucleon size \cite{rho85}.}, namely
\begin{equation}
\frac{m_\rho^*}{m_\rho}=\frac{m_\omega^*}{m_\omega}=\frac{\Lambda^*}{\Lambda}
= 1-0.15\frac{n}{n_0},
\end{equation}
the modifications to correlated 2$\pi$ exchange form the basis of the in-medium
Bonn-B potential \cite{rapp99}. The potential includes also the exchange of the 
$\pi$, $\eta$, and $a_0$ mesons as described in ref.\ \cite{machleidt}. 
Using this model of the medium-modified (MM) Bonn-B potential, the properties of 
symmetric nuclear matter were studied \cite{rapp99}, and it was found that medium effects in
scalar-isoscalar exchange diminished with increasing density, thereby allowing for a 
reasonable description of nuclear saturation in a Dirac-Brueckner-Hartree-Fock calculation.

\begin{figure}[hbt]
\begin{center}
\includegraphics[height=15cm,angle=-90]{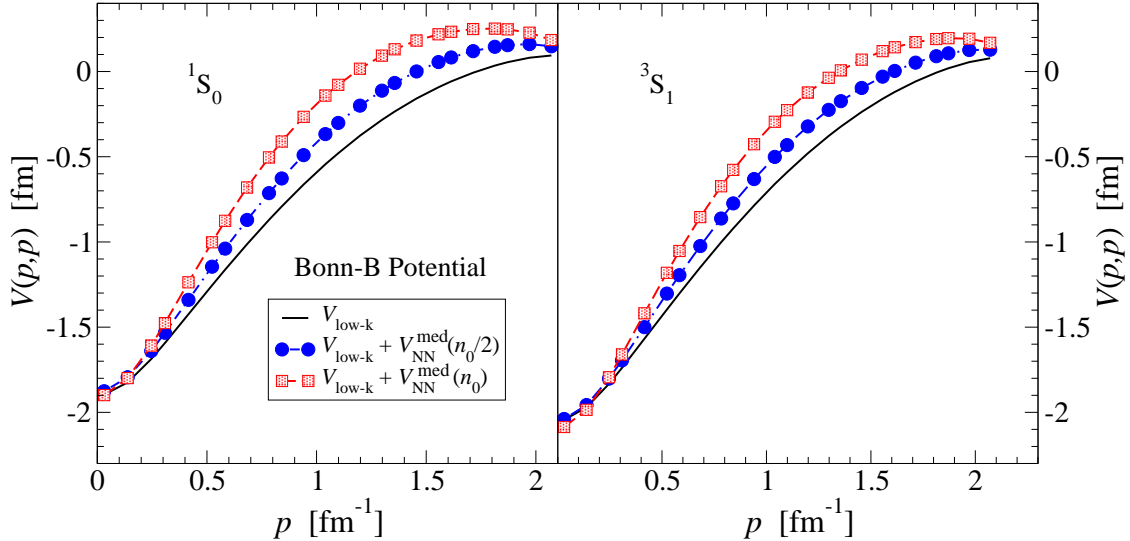}
\caption{Diagonal matrix elements for relative $S$ waves obtained from the low-momentum
medium-modified Bonn-B potential at three densities: $n=\{0,n_0/2,n_0\}$.}
\label{rappbonnpw}
\end{center}
\end{figure}

In Fig.\ \ref{rappbonnpw} we show the on-shell matrix elements of the low-momentum MM
Bonn-B potential in relative $S$-waves for three different densities: $n=\{0,n_0/2$ and $n_0$\}. 
In comparison to the $S$-wave matrix elements found for the density-dependent chiral 
interaction, we see that the main difference lies in the low-momentum components of the 
interaction. Whereas the Pauli-blocked two-pion exchange diagram in Fig.\ \ref{mfig2}
gives rise to a large repulsion at low momenta (see Fig. \ref{n3lopw}), 
the dropping of the effective $\sigma_1$ and
$\sigma_2$ masses in the MM Bonn-B potential increases the medium- and long-range attraction.
Indeed, at nuclear matter saturation density, the effective $\sigma$ meson masses have
dropped to 395 and 220 MeV (from 640 and 377 MeV respectively), 
which extends the range of the attraction significantly.
This enhanced attraction is partially mitigated by a simultaneous drop in the effective coupling
constants $g_{\sigma_1}$ and $g_{\sigma_2}$. However, at higher momenta we see that
the two theories lead to similar modifications in the $S$-wave interactions. In the 
density-dependent chiral interaction the short-range contact interaction proportional
to $c_E$ provides most of the additional repulsion, whereas in the MM Bonn-B potential
an analogous effect is achieved through the dropping of the $\omega$ meson mass.

One of the original motivations for studying the beta decay of $^{14}$C within
the framework of Brown-Rho scaling was the sensitivity of the Gamow-Teller matrix
element to the nuclear tensor force. In fact, employing a residual interaction 
consisting of only central and 
spin-orbit forces it is not possible to achieve a vanishing matrix element in a 
pure two $p$-hole configuration \cite{inglis}. Jancovici and Talmi \cite{talmi} showed 
that by including a strong tensor force one could construct an interaction which 
reproduces the lifetime of $^{14}$C as well as the magnetic moment and electric 
quadrupole moment of $^{14}$N, although agreement with the known spectroscopic data 
was unsatisfactory.
The most important contributions to the tensor component
of the nuclear potential arise from $\pi$ and $\rho$ meson exchange, which have 
opposite signs:
\begin{eqnarray}
V_\rho^T(r) &=& -\frac{f_{N\rho}^2}{4\pi}m_\rho \vec{\tau}_1 \cdot \vec{\tau}_2
 \, S_{12} \left[ \frac{1}{(m_\rho r)^3} + \frac{1}{(m_\rho r)^2}
+ \frac{1}{3 m_\rho r} \right] e^{-m_\rho r},\nonumber \\
V_\pi^T(r) &=& \frac{f_{N\pi}^2}{4\pi}m_\pi \vec{\tau}_1 \cdot \vec{\tau}_2 \, S_{12} 
  \left[ \frac{1}{(m_\pi r)^3} + \frac{1}{(m_\pi
      r)^2}+ \frac{1}{3 m_\pi r} \right]e^{-m_\pi r},
\label{tenforce}
\end{eqnarray}
where $S_{12}=3 \vec \sigma_1 \cdot \hat r\, \vec \sigma_2 \cdot 
\hat r -\vec \sigma_1 \cdot \vec \sigma_2$ is the tensor operator and $f_{N\pi}
=g_A m_\pi / 2f_\pi$.
Since the $\rho$ meson mass is expected to decrease substantially at nuclear
matter density while the pion mass remains relatively constant, an
unambiguous prediction of Brown-Rho scaling is the decreasing of the total tensor force at
finite density. However, as we show in the next section, the additional short-range
repulsive force resulting from the decreasing $\omega$ meson mass also plays an important 
role in suppressing the associated Gamow-Teller strength.

\section{Beta decay lifetime of $^{14}$C}
\label{results}

The extremely long half-life of $^{14}$C ($T_{1/2} \simeq 5730$ yr \cite{ajzen})
plays an
essential role in archaeological dating methods, yet naively one would 
expect the lifetime to be nearly six orders of magnitude smaller. 
The $0^+$ ground state of
$^{14}$C and the $1^+$ ground state of $^{14}$N to which it decays satisfy the
selection for allowed Gamow-Teller transitions, and according to the Wigner 
spin-isospin SU(4) symmetry for light nuclei the associated transition matrix
element would be unity. However, in order to reproduce the experimentally 
observed lifetime, the Gamow-Teller transition matrix element must be
approximately 2$\cdot 10^{-3}$, which is expected to result from an accidental
cancellation among otherwise large terms. This transition therefore provides a
sensitive test for nuclear interaction models and many-body methods.

\begin{figure}[hbt]
\begin{center}
\includegraphics[height=8.3cm,angle=-90]{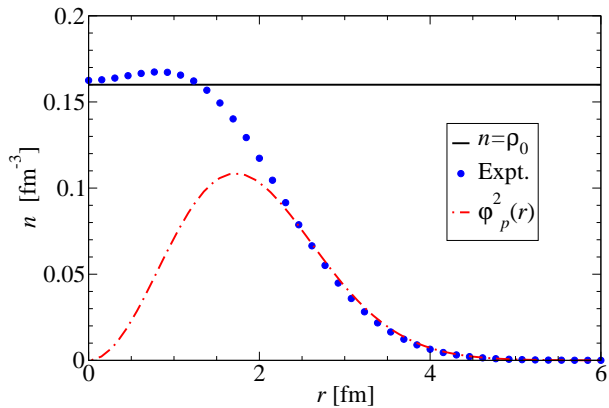}
\caption{Twice the charge distribution of $^{14}$N taken from 
\cite{schaller, schutz} together with the square of the radial $0p$-shell wavefunctions.}
\label{hosc}
\end{center}
\end{figure}

Early studies with phenomenological interactions highlighted the importance
of the nuclear tensor force, without which it would not be possible to obtain
a vanishing Gamow-Teller matrix element in a $0p^{-2}$ model space 
\cite{inglis,talmi}. Employing realistic NN interaction models did not improve
the theoretical description significantly \cite{zamick66}, but the use of larger 
model spaces in no-core shell model calculations appeared to generally diminish
the Gamow-Teller matrix element \cite{aroua}. Despite the long history 
of the problem, it was only recently suggested that 
the beta decay lifetime is particularly sensitive to the density-dependence 
of the nuclear force \cite{holt08,holt09}. The first of these studies employed
the medium modified Bonn-B potential, described in the previous section,
while the latter made use of the density-dependent chiral NN interaction constructed
in Section \ref{ceftnnint}. Indeed, since $^{14}$C and $^{14}$N lie just below the
double shell-closure at $^{16}$O, the density in the vicinity of the valence nucleons
is fairly large. In Fig.\ \ref{hosc} we plot twice the charge distribution of $^{14}$N
obtained from electron scattering experiments \cite{schaller,schutz} fit
to the harmonic oscillator density distribution
\be
n(r) \propto \left(1 + b\frac{r^2}{d^2} \right) e^{-r^2/d^2},
\ee
where for $^{14}$N, $b = 1.29$ and $d=1.74$ fm. Also plotted in the same figure
is the square of the radial $0p$ wavefunction which peaks at a radius where
the nucleon density is 
close to 90\% that of saturated nuclear matter $n_0=0.16$ fm$^{-3}$.
The rest of this section is devoted to a systematic comparison between the
two models of the in-medium nuclear interaction, 
which despite several qualitative differences in their density dependence,
both predict that the beta decay transition of $^{14}$C is strongly suppressed by
medium effects in the nuclear force. 

\begin{figure}[t]
\centering
\includegraphics[height=6cm]{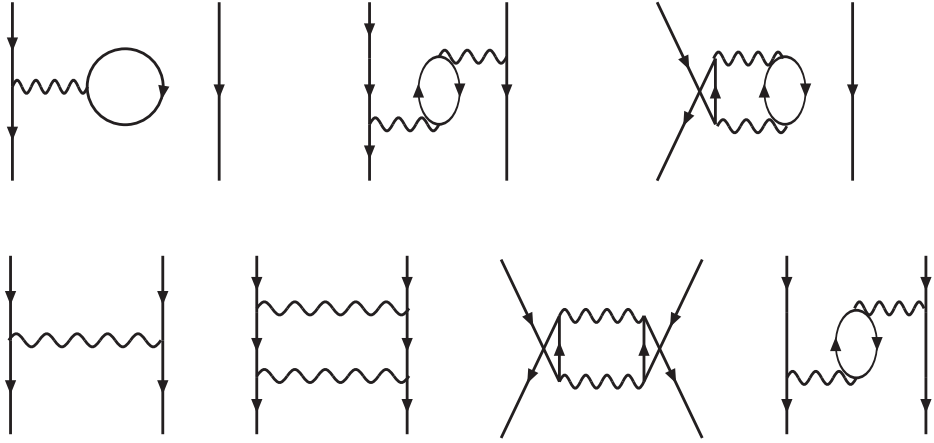}
\caption{Diagrams contributing to the effective interaction $V_{\rm eff}$ in our present calculation. 
Wavy lines represent the density-dependent in-medium nuclear interaction caculated in Sections \ref{ceftnnint}
and \ref{brsnnint}.}
\label{qbox}
\end{figure}

Nuclei with mass number $A>13$ are currently beyond the reach of ab-initio methods, 
such as the no-core shell model \cite{navratil}
and quantum Monte Carlo techniques \cite{pieper}. We approach this problem from shell 
model perturbation theory using a model space consisting of two $0p$ holes in a closed
$^{16}$O core. The many-body Schr\"odinger equation
\be
H\Psi _n=E_n \Psi _n,
\ee
is replaced by the model space equation
\be
H_{\rm eff}\chi _m=E_m \chi _m,
\ee
where $H_{\rm eff}=H_0+V_{\rm eff}$ and $H_0$ is the sum of the kinetic energy and the
harmonic oscillator single particle potential with oscillator parameter 
$\hbar \omega = 14$ MeV. We derive the shell 
model effective interaction $V_{\rm eff}$ following the formalism explained 
in refs.\ \cite{ko90,jensen95}. The effective interaction is constructed from a 
folded diagram expansion in the so-called $\hat Q$-box, which consists of all 
irreducible valence-linked diagrams (usually truncated to a given order). 
In the present calculation we include hole-hole
diagrams in the $\hat Q$-box up to second-order in perturbation theory, shown 
in Fig.\ \ref{qbox}. In this figure the wavy line represents the full 
interaction, either \vlk $+V_{NN}^{\rm med}$ or the low-momentum 
medium-modified Bonn-B potential, $V^{\rm med}_{\rm Bonn-B}$. 
Finally, we need the energy splitting between the two $0p$ orbitals, $\epsilon = e(p_{1/2})- e(p_{3/2})$, 
which is most accurately obtained from the experimental excitation energy of the 
first $3/2^-$ state of $^{15}$N located 6.3 MeV above the $1/2^-$ ground state.

We denote the relevant single-hole states by
\begin{equation}
\left | 1 \right > = \left | 0s_{1/2}^{-1} \right >, 
\left | 2 \right > = \left | 0p_{3/2}^{-1} \right >,
\left | 3 \right > = \left | 0p_{1/2}^{-1} \right >,
\end{equation}
and the two-hole states coupled to good angular momentum $J$, isospin $T$, and 
parity $\pi$ are denoted by $\left | \alpha \beta; J^\pi T \right >$, where 
$\alpha,\beta = $ 2 or 3. After constructing the effective interaction with 
the folded diagram method, we diagonalize the following matrix to obtain the 
eigenvalues and eigenstates with $J^\pi = 1^+, T=0$ (relevant for $^{14}$N):
\begin{equation}
\left[ \begin{array}{ccc}
 & \vdots & \\
\cdots & \langle \alpha \beta; 1^+ 0|V_{\rm eff}|\gamma \delta; 1^+ 0 \rangle & \cdots\\
 & \vdots & \end{array} \right] 
+ \left[\begin{array}{ccc}
0 & 0 & 0 \\
0 & -\epsilon & 0 \\
0 & 0 & -2\epsilon \end{array} \right],
\label{matrices}
\end{equation}
For the $J^\pi = 0^+$, $T=1$ states, we diagonalize a similar 
$2\times 2$ matrix. The ground states for $^{14}$C and $^{14}$N
are the lowest $J^\pi = 0^+$, $T=1$ and $J^\pi = 1^+, T=0$ states respectively,
which we denote by
\begin{eqnarray}
\psi_i &=& a \left |22;0^+ 1 \right > + b\left|33;0^+ 1\right > \, ,\nonumber \\
\psi_f &=& x \left |22;1^+ 0\right > + y\left|23;1^+ 0\right > + 
z\left|33;1^+ 0\right> \, .
\label{ls}
\end{eqnarray}
From these wavefunctions one can derive the following expression for the reduced Gamow-Teller 
matrix element $M_{\rm GT}$:
\begin{equation}
M_{GT} = \sum_{k} \left<\psi_f||\sigma(k)\tau_+(k)||\psi_i\right> = 
\frac{1}{\sqrt{6}}\left(-2\sqrt{5}ax + 2\sqrt{2}ay + 4by + 2bz \right ),
\label{gte}
\end{equation}
where experimentally $M_{GT} \simeq 2 \cdot 10^{-3}$.

For orientation we first determine the low-lying states of 
$^{14}$C and $^{14}$N, as well as the Gamow-Teller transition matrix element
connecting the two ground states,
using the free-space low-momentum N$^3$LO potential at a resolution
scale of $\Lambda_{\rm low-k}=2.1$ fm$^{-1}$. We will see shortly that 
there is only a small difference when using the free-space Bonn-B potential. 
For the two $0^+$ states of $^{14}$C we find
\begin{eqnarray}
\psi_1(0^+, 1) &=& 0.396 \left |22;0^+,1\right > + 
0.918 \left |33;0^+,1\right > \, ,\nonumber \\
\psi_2(0^+,1) &=& 0.918 \left |22;0^+,1\right > - 
0.396 \left |33;0^+,1\right > \, ,
\label{c14wfn}
\end{eqnarray}
with energy splitting $\Delta E (0^+) = 12.9$ MeV. 
For the two lowest $1^+$ states we find
\begin{eqnarray}
\psi_1(1^+, 0) &=& 0.142 \left |22;0^+,1\right > - 
0.681 \left |23;0^+,1\right > + 0.719 \left |33;0^+,1\right > \, ,\nonumber \\
\psi_2(1^+,0) &=& 0.361 \left |22;0^+,1\right > - 
0.663 \left |23;0^+,1\right > - 0.656 \left |33;0^+,1\right >,
\label{n14wfn}
\end{eqnarray}
where the first excited state lies $\Delta E (1^+) = 2.76$ MeV above the ground state. 
The second excited $J^\pi =1^+, T=0$ state lies nearly 16 MeV higher in energy than the 
ground state and will therefore be neglected throughout. With these wavefunctions, the 
ground state to ground state Gamow-Teller transition matrix element is found to be
\begin{equation}
M_{GT} = -0.90\, ,
\label{gtv}
\end{equation}
which is much too large to describe the known lifetime of $^{14}$C. From the ground
state wavefunctions in eqs.\ (\ref{c14wfn}) and (\ref{n14wfn}) we see that the 
first three terms in eq.\ (\ref{gte}) add coherently
while the last gives a contribution of the opposite sign, though not enough to cancel
the first three. In order for medium effects to suppress the transition, strength
must therefore be shifted to the $|33\rangle$ component of both ground state
wavefunctions.

We now implement density-dependence in the nuclear potential and study the
effect on the model space interaction matrix elements. 
In Tables \ref{tnf0cb} and \ref{tnf1cb} we show the shell model matrix elements for the 
low-momentum interactions derived from the in-medium chiral N$^3$LO and Brown-Rho-scaled (BRS)
Bonn-B potentials both in free space and at nuclear matter saturation density. 
Table \ref{tnf0cb} shows the matrix elements for the 
$J^\pi = 0^+, T=1$ states, while Table \ref{tnf1cb} shows the
matrix elements for the $J^\pi = 1^+, T=0$ states. Overall, medium effects in
both potentials lead to qualitatively similar changes in the shell model matrix
elements, though the density-dependence of the in-medium chiral interaction
is much stronger. This should not be surprising, given the stronger $S$-wave 
repulsion in the in-medium chiral interaction, as compared to the BRS Bonn-B 
potential, shown in Figs.\ \ref{swaves} and \ref{rappbonnpw}.

\setlength{\tabcolsep}{.075in}
\begin{table}[htb]
\begin{center}
\begin{tabular}{|c|c|c|c|} 
\multicolumn{4}{c}{$J^\pi = 0^+, T=1$} \\ \hline
$n/n_0$ & $\langle 22| V_{\rm Chiral}^{\rm med}|22\rangle$ & $\langle22|
V_{\rm Chiral}^{\rm med} |33\rangle$ 
& $\langle33|V_{\rm Chiral}^{\rm med} |33\rangle$  \\ \hline
0 &  -3.28 &  -3.77 &  -0.61  \\ \hline
0.25 & -2.64 & -3.66 & -0.05 \\ \hline
1.0 &  -0.12 & -2.88 & 1.91 \\ \hline
\multicolumn{4}{l}{} \\ \hline
$n/n_0$ & $\langle 22| V_{\rm Bonn}^{\rm med}|22\rangle$ & $\langle22|
V_{\rm Bonn}^{\rm med} |33\rangle$ 
& $\langle33|V_{\rm Bonn}^{\rm med} |33\rangle$  \\ \hline
0 & -3.12 &  -3.65 &  -0.54  \\ \hline
0.25 & -2.72 &  -3.38 &  -0.33  \\ \hline
1.0 & -1.09 &  -2.21 &  0.48  \\ \hline
\end{tabular}
\caption{Matrix elements (in units of MeV) between $0p^{-2}$ states coupled to 
$(J^\pi,T) = (0^+,1)$ for the in-medium chiral interaction as well
as the Brown-Rho-scaled Bonn-B potential 
at three different densities: $n=0, 0.25n_0,$ and $n_0$.}
\label{tnf0cb}
\end{center}
\end{table}

\setlength{\tabcolsep}{.025in}
\begin{table}[htb]
\begin{center}
\begin{tabular}{|c|c|c|c|c|c|c|} 
\multicolumn{7}{c}{$J^\pi = 1^+, T=0$} \\ \hline
$n/n_0$ & $\langle 22| V_{\rm Chiral}^{\rm med}|22 \rangle$ & $\langle 23|
V_{\rm Chiral}^{\rm med} |22\rangle$ & 
$\langle 23|V_{\rm Chiral}^{\rm med} |23\rangle$ & $\langle 23|
V_{\rm Chiral}^{\rm med} |33\rangle$ & 
$\langle 22|V_{\rm Chiral}^{\rm med} |33\rangle$ & $\langle 33|
V_{\rm Chiral}^{\rm med} |33\rangle$  \\ \hline
0.0 & -1.19 & 3.89 & -5.20 & 1.36 & 1.66 & -1.67 \\ \hline
0.25 & -0.61 & 3.74 & -4.81 & 1.50 & 1.33 & -1.00 \\ \hline
1.0 & 1.29 & 2.81 & -3.03 & 1.92 & 0.44 & 0.72 \\ \hline
\multicolumn{7}{l}{} \\ \hline
$n/n_0$ & $\langle 22| V^{\rm med}_{\rm Bonn} |22\rangle$ & $\langle 23|
V^{\rm med}_{\rm Bonn} |22\rangle $ & 
$\langle 23|V^{\rm med}_{\rm Bonn} |23\rangle$ & $\langle 23|
V^{\rm med}_{\rm Bonn} |33\rangle$ & 
$\langle 22|V^{\rm med}_{\rm Bonn} |33\rangle$ & 
$\langle 33|V^{\rm med}_{\rm Bonn} |33\rangle$  \\ \hline
0.0 &  -1.28 & 3.50 & -4.91 & 1.23 & 1.46 & -1.68 \\ \hline
0.25 & -1.09 & 3.27 & -4.56 & 1.24 & 1.38 & -1.57 \\ \hline
1.0 &  -0.31 & 2.31 & -3.11 & 1.26 & 1.04 & -1.09 \\ \hline
\end{tabular}
\caption{Matrix elements (in units of MeV) between $0p^{-2}$ states coupled to 
$(J^\pi,T) = (1^+,0)$ for the in-medium chiral interaction as well
as the Brown-Rho-scaled Bonn-B potential 
at three different densities: $n=0, 0.25n_0,$ and $n_0$.}
\label{tnf1cb}
\end{center}
\end{table}

At nuclear matter saturation density, medium effects on both potentials are quite
large. Nevertheless, we can gain insight by studying the interactions at low densities 
and employing perturbation theory. 
One finds the following corrections to the ground state wavefunctions:
\begin{eqnarray}
 \psi_1^{(1)}(0^+,1) &=& \psi_1^{(0)}(0^+,1) - \frac{1}{12.9} \left [
0.36\langle 22 | V^{\rm med} | 22\rangle +0.69 \langle 22 | V^{\rm med} | 33\rangle \right .
\nonumber \\
&-& \left . 0.36 \langle 33 | V^{\rm med} | 33\rangle \right ] \psi_2^{(0)}(0^+,1)
\label{c14gsp}
\end{eqnarray}
and
\begin{eqnarray}
 \psi_1^{(1)}(1^+,0) &=& \psi_1^{(0)}(1^+,0) - \frac{1}{2.76} \left [
0.05\langle 22 | V^{\rm med} | 22\rangle -0.34 \langle 23 | V^{\rm med} | 22\rangle
+ 0.45 \langle 23 | V^{\rm med} | 23\rangle \right . \nonumber \\
&-& \left . 0.03 \langle 23 | V^{\rm med} | 33\rangle
+ 0.17 \langle 22 | V^{\rm med} | 33\rangle 
 -0.47 \langle 33 | V^{\rm med} | 33\rangle 
\right ] \psi_2^{(0)}(1^+,0).
\label{n14gsp}
\end{eqnarray}

A straightforward calculation using the matrix elements shown in 
Tables \ref{tnf0cb} and \ref{tnf1cb} for $n = 0.25 n_0$ gives for 
the un-normalized perturbed wavefunctions arising from 
$V_{\rm Chiral}^{\rm med}$
\begin{eqnarray}
\psi^{(1)}_0(0^+,1) &=& \psi^{(0)}_0(0^+,1) - 0.01 \psi^{(0)}_1(0^+,1)\, \hspace{.1in} {\rm and} \nonumber \\
\psi^{(1)}_0(1^+,0) &=& \psi^{(0)}_0(1^+,0) + 0.04 \psi^{(0)}_1(1^+,0)\, ,
\label{pert25}
\end{eqnarray}
We see that a positive $\langle 22; 0^+ 1 | V^{\rm med} | 22; 0^+ 1 \rangle$ 
and a positive $\langle 22; 0^+ 1 | V^{\rm med} | 33; 0^+ 1 \rangle$ lead to a 
stronger $|33; 0^+,1\rangle$ component in the perturbed $^{14}$C ground state 
wavefunction. One therefore expects quite generally that additional 
density-dependent repulsion would increase the $|33; 0^+ 1\rangle$ component 
in the $^{14}$C ground state and therefore help suppress the beta decay of 
$^{14}$C. Both the in-medium chiral nuclear interaction and the BRS Bonn-B
potential share this feature. 

For the ground state of $^{14}$N the situation is
more complicated due to the larger number of terms in the perturbed wavefunction of
eq.\ (\ref{n14gsp}) as well as the more complicated dependence of the shell model 
matrix elements on the perturbing potential. The two dominant terms are
$\langle 33; 1^+ 0 | V^{\rm med} | 33; 1^+ 0 \rangle$ and 
$\langle 23; 1^+ 0 | V^{\rm med} | 23; 1^+ 0 \rangle$, which have the same sign as 
the perturbing potential in $S$-waves but enter with opposite sign in the perturbed 
wavefunction. At one-quarter of nuclear matter density, the sum of all six
contributions to $V_{\rm Chiral}^{\rm med}$ yields 
$\langle 33; 1^+ 0 | V_{\rm Chiral}^{\rm med} | 33; 1^+ 0 \rangle= 0.67$ MeV and
$\langle 23; 1^+ 0 | V_{\rm Chiral}^{\rm med} | 23; 1^+ 0 \rangle=0.39$ MeV. The
remaining four terms together give a minute contribution to the perturbed 
wavefunction, and therefore we expect at low densities for the $|33; 1^+,0\rangle$
component in the ground state of $^{14}$N to decrease for the in-medium
chiral interaction and consequently increase the magnitude of the 
Gamow-Teller transition matrix element. Since different components of the
in-medium chiral interaction have 
different density-dependences, this behavior can change at higher 
densities. Indeed by inspecting the shell model matrix elements at 
nuclear matter saturation density we see that the opposite effect occurs and
the GT matrix element is suppressed rather than enhanced (the 
perturbing potential is quite large and therefore our argument based on 
perturbation theory should be treated cautiously).
In contrast, the BRS Bonn-B potential has
$\langle 33; 1^+ 0 | V_{\rm Bonn}^{\rm med} | 33; 1^+ 0 \rangle= 0.11$ MeV and
$\langle 23; 1^+ 0 | V_{\rm Bonn}^{\rm med} | 23; 1^+ 0 \rangle=0.35$ MeV, which
suggests that even at low densities the beta decay should be suppressed.

We now consider the results of the full calculation, including all density-dependent
contributions to second-order in the shell model effective 
interaction shown in Fig.\ \ref{qbox}. In Table \ref{n3locoef} we show the results of our 
calculations for the 
expansion coefficients of the ground state wavefunctions of $^{14}$C and $^{14}
$N, as well as the reduced GT matrix element, as a function of the nuclear 
density up to $n = n_0$.
\setlength{\tabcolsep}{.075in}
\begin{table}[t]
\begin{center}
\begin{tabular}{|c||c|c|c|c|c|c||c|c|c|c|c|c|} \hline
 & \multicolumn{6}{|c||}{In-medium chiral N$^3$LO} & \multicolumn{6}{|c|}{In-medium Bonn-B} \\ \hline
$n/n_0$ & $a$ & $b$ & $x$ & $y$ & $z$ & $M_{\rm GT}$ 
& $a$ & $b$ & $x$ & $y$ & $z$ & $M_{\rm GT}$ \\ \hline
0.0 & 0.40 & 0.92 & 0.14 & -0.68 & 0.72 & -0.90
& 0.40 & 0.93 & 0.11 & -0.62 & 0.78 & -0.70 \\ \hline
0.2 & 0.38 & 0.92 & 0.15 & -0.69 & 0.71 & -0.91
& 0.35 & 0.94 & 0.08 & -0.55 & 0.83 & -0.49 \\ \hline
0.4 & 0.36 & 0.93 & 0.14 & -0.67 & 0.73 & -0.82 
& 0.32 & 0.95 & 0.05 & -0.50 & 0.86 & -0.32 \\ \hline
0.6 & 0.33 & 0.94 & 0.13 & -0.65 & 0.75 & -0.74
& 0.29 & 0.96 & 0.03 & -0.45 & 0.89 & -0.17 \\ \hline
0.8 & 0.30 & 0.96 & 0.12 & -0.62 & 0.78 & -0.64
& 0.25 & 0.97 & 0.02 & -0.40 & 0.92 & -0.04 \\ \hline
1.0 & 0.26 & 0.97 & 0.11 & -0.59 & 0.80 & -0.53
& 0.21 & 0.98 & 0.01 & -0.36 & 0.93 &  0.07 \\ \hline
\end{tabular}
\caption{The coefficients of the $jj$-coupled wavefunctions defined in eq.\
  (\ref{ls}) and the associated reduced GT matrix element as a function of the nuclear
  density $n$ for both the Brown-Rho-scaled Bonn-B potential and the in-medium
chiral nuclear interaction.}
\label{n3locoef}
\end{center}
\end{table}
The full results agree well with the conclusions we obtained from arguments based on perturbation theory. 
The Gamow-Teller transition matrix element decreases 
immediately at low densities in the BRS Bonn-B potential, while only after 
$n\simeq0.25n_0$ do the density-dependent corrections of the chiral nuclear interaction
lead to a suppression of the matrix element.

The Gamow-Teller transition strength, $B(GT)$, is related to the reduced GT transition matrix element by 
\begin{equation}
B(GT)=(g_A^*)^2 \frac{1}{2J_i+1}|M_{\rm GT}|^2 \simeq |M_{\rm GT}|^2,
\end{equation}
where $J_i =0$ and we have used the approximation that the in-medium 
axial vector coupling constant $g_A^* \simeq 1$. In general, the effective Gamow-Teller operator has 
additional terms in a nuclear medium \cite{arima,castel}:
\begin{equation}
{\cal \vec{O}}_{\rm GT,eff} = g_{LA} \vec{L} + g_A^* \, \vec{\sigma} + g_{PA} \,[Y_2,\vec{\sigma}],
\end{equation}
where $\vec{L}$ is the orbital angular momentum operator and $Y_2$ denotes the rank-2 
spherical harmonic $Y_{2,m}$. However, from theoretical calculations \cite{towner} of the
effective GT operator in-medium, as well as beta decay calculations \cite{babrown} performed 
with phenomenological shell model effective interations, it is known that $g_{LA}$ and $g_{PA}$ are almost 
negligible and that in light nuclei $g_A^*$ is smaller by 15-20\% 
from its free space value of $g_A = 1.27$ (measured in free neutron beta decay). 
In general, one 
must calculate both the effective interaction $V_{\rm eff}$ and the effective Gamow-Teller 
operator ${\cal \vec{O}}_{\rm GT, eff}$. In our calculations we assume a 20\%
reduction of $g_A$ in medium and therefore set $g_A^* = 1.0$.

\begin{figure}[thb]
\centering
\includegraphics[height=17cm, angle=-90]{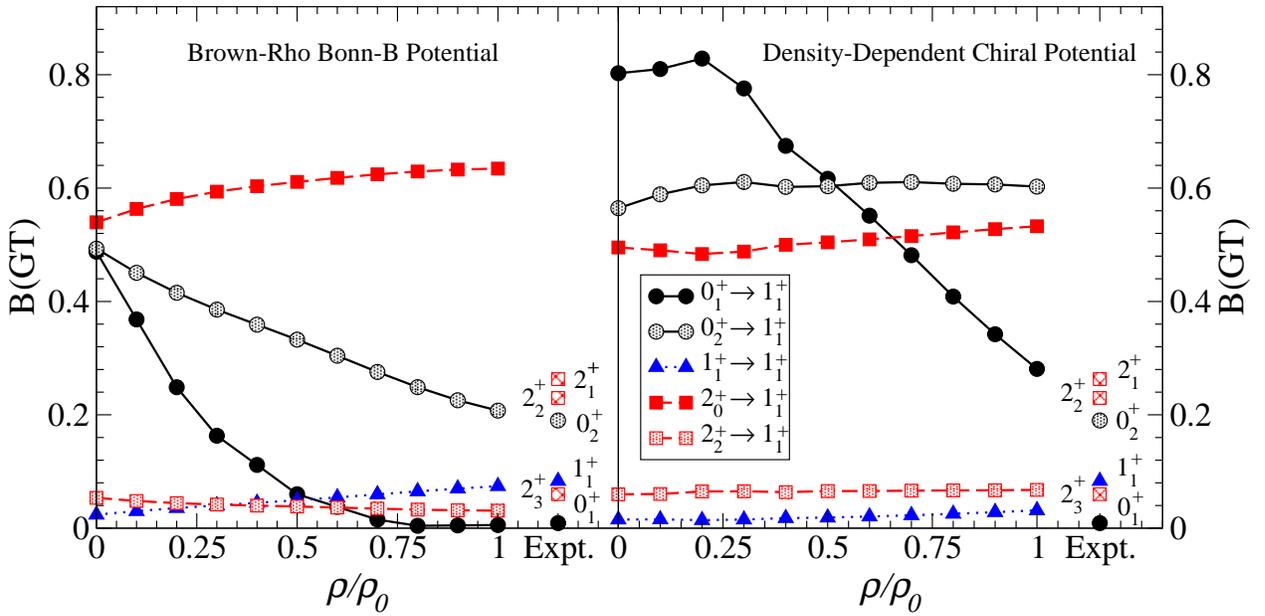}
\caption{The $B(GT)$ values for transitions from low-lying states 
of $^{14}$C to the ground state of $^{14}$N
 as a function of the nuclear density. The
  experimental values are from \cite{negret}. Note that there are three
  experimental low-lying $2^+$ states compared to two theoretical $2^+$ states
  in the $0p^{-2}$ shell model configuration.}
\label{bgtn3lo}
\end{figure}

In Fig.\ \ref{bgtn3lo} we plot the $B(GT)$ strengths for transitions from the low-lying
states of $^{14}$C to the ground state of $^{14}$N for both density-dependent
NN interactions. These are compared to the experimental values \cite{negret} obtained from the 
charge-exchange reaction $^{14}{\rm N}(d, {^{2}{\rm He}})^{14}{\rm C}$.
From Fig.\ \ref{bgtn3lo} one sees that the most dramatic effect is a reduction in the ground
state to ground state Gamow-Teller transition strength in both models. In the 
Brown-Rho-scaled Bonn-B potential also the first excited ($0^+,1$) state of $^{14}$C
is decreased toward the experimental value. Such an effect can be obtained from
in-medium chiral interactions, but only if the strength of the contact three-body
force is increased \cite{holt09}. In the present version of the in-medium chiral nuclear
interaction, there appears to be very little effect on the other Gamow-Teller 
transition strengths.

\begin{figure}[t]
\begin{center}
\includegraphics[height=13cm, angle=-90]{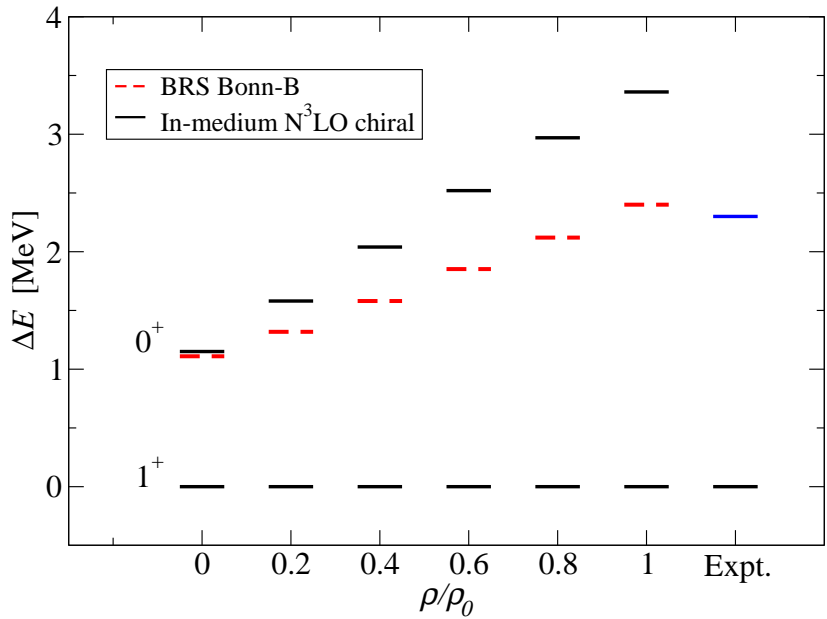}
\caption{The splitting between the 1$^+$ ground state of $^{14}$N and the first excited 
0$^+$ state as a function of the nuclear density for both the medium-modified Bonn-B potential
and the in-medium chiral nuclear interaction. Also shown is the experimental energy splitting 
$\Delta E_{0^+-1^+} = 2.3$ MeV.}
\label{elevel}
\end{center}
\end{figure}

Besides comparing the Gamow-Teller strengths within the two models, 
it is useful to study the energy splitting between the ground state of 
$^{14}$N and the first excited ($0^+,1$) state (the isobaric analogue 
of the $^{14}$C ground state), which is experimentally observed to lie
at 2.3 MeV. In Fig.\ \ref{elevel} we plot
this energy splitting as a function of the nuclear density for the in-medium
chiral nuclear interaction and the BRS Bonn-B potential. It is clear that 
free-space nucleon-nucleon interactions systematically underpredict this level
splitting and that additional density-dependent repulsion leads to an enhancement
of the splitting. The in-medium chiral nuclear interaction is much more
repulsive than the BRS Bonn-B potential and therefore leads to a larger increase in
the splitting which overpredicts the value at higher densities. Nevertheless, we find
that qualitatively the two models for the density-dependent NN interaction are consistent
with one another and improve the agreement with experiment.

In order to better understand the relationship between the in-medium chiral 
nuclear interaction and
the BRS Bonn-B potential, we first study in detail the six different one-loop contributions 
to $V_{NN}^{\rm med}$ arising from the leading-order chiral three-nucleon force. In Tables 
\ref{tnf0c} and \ref{tnf1c} we show the most important shell model matrix elements for
each of these components at nuclear matter saturation density $n_0$. Again, we let 
perturbation theory arguments direct our attention to the relevant physics. As previously
mentioned, in order to suppress the Gamow-Teller transition matrix element, it is necessary
to increase the $|33; J^\pi T \rangle$ components of the ground state wavefunctions for
both $^{14}$C and $^{14}$N. From eq.\ (\ref{c14gsp}) it follows that only 
$V^{\rm med,1}_{\rm NN}$, $V^{\rm med,3}_{\rm NN}$, and $V^{\rm med,6}_{\rm NN}$
contribute in the desired way. The remaining three contributions to $V_{\rm NN}^{\rm med}$
are attractive and decrease the strength of the $|33;0^+,1\rangle$ component of the $^{14}$C
ground state wavefunction. The structure of the $^{14}$N perturbed ground state is most 
strongly affected by the $\langle 33; 1^+ 0 | V^{\rm med} | 33; 1^+ 0 \rangle$ and 
$\langle 23; 1^+ 0 | V^{\rm med} | 23; 1^+ 0 \rangle$ matrix elements. A secondary
role is played by the $\langle 22; 1^+ 0 | V^{\rm med} | 23; 1^+ 0 \rangle$ and 
$\langle 22; 1^+ 0 | V^{\rm med} | 33; 1^+ 0 \rangle$ components. From Table \ref{tnf1c}
we see that out of the three contributions that shift strength to the $|33; 0^+ 1 \rangle$ 
component of the $^{14}$C ground state, only the three-nucleon contact interaction 
proportional to $c_E$ also strongly shifts strength to the $|33; 1^+ 0 \rangle$ component
of $^{14}$N:
\begin{eqnarray}
V_{\rm NN}^{\rm med,1}&:& \psi^{(1)}_0(1^+,0) = \psi^{(0)}_0(0^+,1) + 0.25 \psi^{(0)}_1(0^+,1)\, , \nonumber \\
V_{\rm NN}^{\rm med,3}&:& \psi^{(1)}_0(1^+,0) = \psi^{(0)}_0(1^+,0) - 0.03 \psi^{(0)}_1(1^+,0)\, , \nonumber \\
V_{\rm NN}^{\rm med,6}&:& \psi^{(1)}_0(1^+,0) = \psi^{(0)}_0(1^+,0) - 0.25 \psi^{(0)}_1(1^+,0)\, .
\label{pertcomp}
\end{eqnarray}
A full treatment shows that $V_{\rm NN}^{\rm med,3}$ alone has little effect on the ground
state wavefunction of $^{14}$N and in fact actually decreases the strength in the 
$|33; 1^+ 0 \rangle$ component of the wavefunction. Therefore, the strong short-distance 
repulsion generated by the three-nucleon contact interaction is the dominant term contributing
to the suppression the Gamow-Teller transition matrix element. 

\setlength{\tabcolsep}{.075in}
\begin{table}[htb]
\begin{center}
\begin{tabular}{|c|c|c|c|} 
\multicolumn{4}{c}{$J^\pi = 0^+, T=1$ \hspace{.2in} ($n=n_0$)} \\ \hline
 & $\langle 22| V_{\rm Chiral}^{\rm med}|22\rangle$ & $\langle22|
V_{\rm Chiral}^{\rm med} |33\rangle$ 
& $\langle33|V_{\rm Chiral}^{\rm med} |33\rangle$  \\ \hline
1 &  3.88 &  0.80 &  3.32  \\ \hline
2 & -3.18 & -0.71 & -2.68  \\ \hline
3 &  2.15 &  0.32 &  1.92  \\ \hline
4 & -0.56 & -0.12 & -0.47  \\ \hline
5 & -0.89 & -0.64 & -0.44  \\ \hline
6 &  1.75 &  1.24 &  0.88  \\ \hline
\end{tabular}
\caption{Matrix elements (in units of MeV) between $0p^{-2}$ states coupled 
to $(J^\pi,T) = (0^+,1)$ for the six density-dependent contributions to the 
in-medium chiral nuclear interaction at nuclear matter saturation density $n_0$.}
\label{tnf0c}
\end{center}
\end{table}
\setlength{\tabcolsep}{.025in}
\begin{table}[htb]
\begin{center}
\begin{tabular}{|c|c|c|c|c|c|c|} 
\multicolumn{7}{c}{$J^\pi = 1^+, T=0$ \hspace{.2in} ($n=n_0$)} \\ \hline
$n/n_0$ & $\langle 22| V_{\rm Chiral}^{\rm med}|22 \rangle$ & $\langle 23|
V_{\rm Chiral}^{\rm med} |22\rangle$ & 
$\langle 23|V_{\rm Chiral}^{\rm med} |23\rangle$ & $\langle 23|
V_{\rm Chiral}^{\rm med} |33\rangle$ & 
$\langle 22|V_{\rm Chiral}^{\rm med} |33\rangle$ & $\langle 33|
V_{\rm Chiral}^{\rm med} |33\rangle$  \\ \hline
1 &  4.25 & -0.65 &  1.67 &  2.14 & -1.67 &  3.28 \\ \hline
2 & -3.15 &  0.73 & -1.47 & -1.44 &  1.47 & -2.65 \\ \hline
3 &  1.36 & -0.76 &  1.53 & 0.00 & -0.86 &  1.70 \\ \hline
4 & -0.54 &  0.13 & -0.27 & -0.23 &  0.25 & -0.47 \\ \hline
5 &  -0.47 & 0.56 &  -1.02 &  0.100 & 0.14 &  -0.35 \\ \hline
6 &  1.04 & -1.09 &  1.73 & -0.01 & -0.55 &  0.88 \\ \hline
\end{tabular}
\caption{Matrix elements (in units of MeV) between $0p^{-2}$ states coupled 
to $(J^\pi,T) = (1^+,0)$ for the six density-dependent contributions to the 
in-medium chiral nuclear interaction at nuclear matter saturation density $n_0$.}
\label{tnf1c}
\end{center}
\end{table}

\begin{figure}[thb]
\begin{center}
\includegraphics[height=15cm, angle=-90]{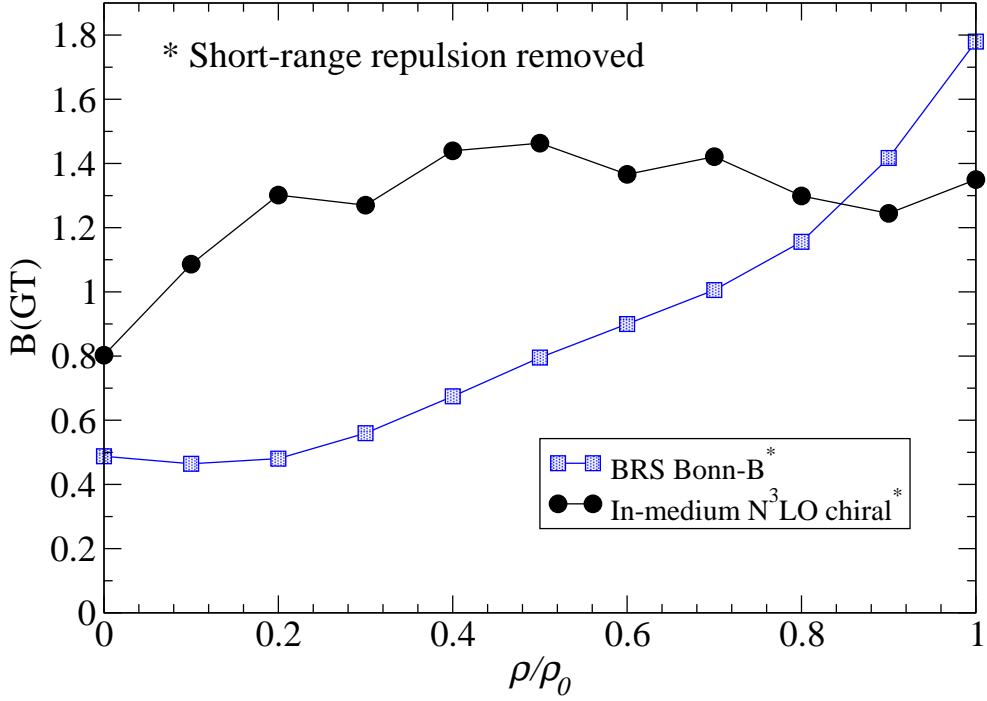}
\caption{The $B(GT)$ values for the $^{14}$C ground state to $^{14}$N ground state
transition as a function of the density for the medium-modified Bonn-B potential and the 
in-medium chiral nucleon-nucleon potential. The asterisk denotes that the additional 
density-dependent short-range repulsion has been removed from both potentials: the $\omega$
meson scaling in the case of the MM Bonn-B potential and the short-range contact three-nucleon
force proportional to $c_E$ in the in-medium chiral interaction.}
\label{bgtn3lowoce}
\end{center}
\end{figure}

In the BRS Bonn-B potential one might expect analogously 
that the scaling of the $\omega$-meson mass plays a similar role. 
We therefore have performed a similar calculation using a BRS Bonn-B
potential with no $\omega$-meson mass scaling. In Fig.\ \ref{bgtn3lowoce}
we have plotted the $B(GT)$ values of the ground state to ground state
transition using medium-modified interactions with the short-distance
repulsion removed (the three-body contact term in the case of the in-medium
chiral interaction and the dropping $\omega$ meson mass in the case of the
BRS Bonn-B potential). We find that in both cases there is in fact a strong
enhancement of the Gamow-Teller transition. Therefore, despite the fact that 
these two density-dependent interactions are behaving very differently at
long distances (as inferred from the low-momentum components of their interactions
in Figs.\ \ref{swaves} and \ref{rappbonnpw}), they both include a strong
short-distance repulsion that dramatically suppresses the Gamow-Teller
transition matrix element.

\section{Summary and Conclusions}
We have studied the effect of density-dependent low-momentum nucleon-nucleon
interactions on the beta decay lifetime of $^{14}$C. The 
in-medium chiral nuclear interaction was obtained from the leading-order
three-nucleon force by contracting one pair of external lines and summing over
the filled Fermi sea of nucleons, whereas the Brown-Rho-scaled Bonn-B
potential was obtained by decreasing the masses and form factor cutoffs of
the vector mesons and treating microscopically effective $\sigma$ meson exchange
as correlated two-pion exchange. Although both density-dependent interactions
introduce additional strong repulsion at short distances, they differ largely at
intermediate and long distances. The Brown-Rho scaled Bonn-B potential
enhances the attraction at these distances (through a decreasing ``$\sigma$'' meson mass), 
but the Pauli-blocked two-pion exchange
diagram obtained from the long-range chiral three-nucleon force is strongly
repulsive in this region. Nevertheless, we find that in both models of the in-medium
nuclear interaction the ground state to ground state Gamow-Teller transition strength
is strongly suppressed at high densities. In contrast, the GT strengths from 
the ground state of $^{14}$N to the excited states of $^{14}$C exhibit only a 
small density dependence. From perturbative considerations 
we have found that, in general, short-range repulsive interactions lead to a
strong suppression of the ground state to ground state transition. Moreover, we have explicitly 
removed the source of short-range repulsion in both interactions and found that
this leads rather to an enhancement of the Gamow-Teller strength. We therefore
suggest that the common short-distance behavior of these two interactions is
responsible for their qualitatively similar effects on the beta decay lifetime 
of $^{14}$C.

\vspace{.1in}

\begin{center}
---------------------------
\end{center}

\noindent The authors of this work represent three generations of former
students and postdocs who have been immensely inspired by Gerry Brown's
physics intuition and ideas. We gratefully dedicate this paper to Gerry
on the occasion of his 85$^{\rm th}$ birthday.

\clearpage

\section{Appendix: Explicit expressions for the Fermi sphere integrals in
 $V_{NN}^{\rm med}$}

In this appendix we tabulate the different functions 
that contribute to the density-dependent expressions in eqs.\ (\ref{med1})-(\ref{med6}). 
The functions $\Gamma_j(p)$ result from Fermi sphere integrals over a single static pion propagator
and have the explicit forms:

\begin{equation} \Gamma_0(p) = k_f - m_\pi \bigg[ \arctan{k_f+p \over m_\pi} +
\arctan{k_f-p \over m_\pi}\bigg] + {m_\pi^2 +k_f^2 -p^2 \over 4p}\ln {m_\pi^2  
+(k_f+p)^2 \over m_\pi^2+(k_f-p)^2} \,, \end{equation}   
\begin{equation} \Gamma_1(p) = {k_f \over 4p^2}  (m_\pi^2+k_f^2+p^2) -\Gamma_0(p)
-{1\over 16 p^3} \Big[m_\pi^2 +(k_f+p)^2\Big]\Big[ m_\pi^2+(k_f-p)^2\Big]\ln 
{m_\pi^2   +(k_f+p)^2 \over m_\pi^2+(k_f-p)^2} \,, \end{equation} 
 \begin{equation} \Gamma_2(p)= {k_f^3 \over 9}+{1\over6} (k_f^2-m_\pi^2-p^2)
\Gamma_0(p)+{1\over6} (m_\pi^2+k_f^2-p^2)\Gamma_1(p)\,, \end{equation}
\begin{equation} \Gamma_3(p)= {k_f^3 \over 3p^2}-{m_\pi^2+k_f^2+p^2 \over  2p^2}
\Gamma_0(p)-{m_\pi^2+k_f^2+3p^2 \over 2p^2}\Gamma_1(p)\,. \end{equation}

\noindent The $G_j(p,q)$ functions result from Fermi sphere integrals over the product of two
static pion propagators. Angular integrations are performed analytically, and the
remaining relevant momentum integrals are given by

\begin{equation} G_{0,*,**}(p,q) = {2\over q} \int_0^{k_f}\!\! dk\,  {\{k,k^3,k^5\} 
\over \sqrt{A(p)+q^2 k^2} } \ln { q\, k+\sqrt{A(p)+q^2 k^2}\over \sqrt{A(p)}}\,,
\end{equation} 
where $A(p)= [m_\pi^2 +(k+p)^2][ m_\pi^2+(k-p)^2]$. The remaining
functions $G_j(p,q)$ are related to those above through the equations
\begin{eqnarray} G_1(p,q)&=& {\Gamma_0(p)-(m_\pi^2+p^2)G_0(p,q) -G_*(p,q) \over 
4p^2-q^2} \,,\\ G_{1*}(p,q)&=&  {3\Gamma_2(p)+p^2\Gamma_3(p)-(m_\pi^2+p^2)G_*(p,q)
  -G_{**}(p,q) \over  4p^2-q^2} \,,\\G_2(p,q)&=&(m_\pi^2+p^2)G_1(p,q)+G_*(p,q)+
G_{1*}(p,q)\,,\\ G_3(p,q)&=& {{1\over 2}\Gamma_1(p)-2(m_\pi^2+p^2)G_1(p,q) 
-2G_{1*}(p,q) -G_{*}(p,q) \over 4p^2-q^2} \,.\end{eqnarray} 
In this chain of equations the functions indexed with an asterisk play 
only an auxiliary role for the construction of $G_{1,2,3}(p,q)$. We note that all 
functions $G_j(p,q)$ are non-singular at $q=2p$ (corresponding to scattering
in backward direction). For notational simplicity, the $k_f$-dependence of $\Gamma_j(p)$
and $G_j(p,q)$ has been suppressed.

\end{document}